\documentclass[12pt,a4paper,english,onecolumn,draftcls]{IEEEtran}
\usepackage[T1]{fontenc}
\usepackage[latin9]{inputenc}
\usepackage{xcolor}
\usepackage{verbatim}
\usepackage{float}
\usepackage{mathrsfs}
\usepackage{amsmath}
\usepackage{amsthm}
\usepackage{amssymb}
\usepackage{graphicx}
\usepackage{setspace}
\makeatletter

\pdfpageheight\paperheight
\pdfpagewidth\paperwidth

\theoremstyle{plain}
\newtheorem{thm}{\protect\theoremname}
\theoremstyle{plain}
\newtheorem{lem}[thm]{\protect\lemmaname}

\usepackage{subfigure}
\usepackage{epstopdf}
\usepackage{cite}
\usepackage{citesort}
\usepackage{balance}

\makeatother

\usepackage{babel}
\providecommand{\lemmaname}{Lemma}
\providecommand{\theoremname}{Theorem}

\begin{document}

\title{% paper title
}

\title{Optimal Base Station Antenna Downtilt in Downlink Cellular Networks}

\author{\noindent \IEEEauthorblockA{\noindent Junnan Yang, Student Member, IEEE, Ming Ding, Senior Member,
IEEE, Guoqiang Mao, Fellow, IEEE, Zihuai Lin, Senior Member, IEEE,
De-gan Zhang, Member, IEEE, Tom Hao Luan, Member, IEEE }% <-this % stops a space
}
\maketitle
\begin{abstract}
From very recent studies, the area spectral efficiency (ASE) performance
of downlink (DL) cellular networks will continuously decrease and
finally to zero with the network densification in a fully loaded ultra-dense
network (UDN) when the absolute height difference betwee\textcolor{black}{n
a ba}se station (BS) antenna and \textcolor{black}{a} user equipment
(UE) antenna is larger than zero, which is referred as the ASE Crash.
We revisit this issue by considering the impact of the BS antenna
downtilt on the downlink network capacity. In general, there exists
a height difference between a BS and a UE in practical networks. It
is common to utilize antenna downtilt to adjust the direction of the
vertical antenna pattern, and thus increase received signal power
or reduce inter-cell interference power to improve network performance.
This paper focuses on investigating the relationship between the base
station antenna downtilt and the downlink network capacity in terms
of the coverage probability and the ASE. The analytical results of
the coverage probability and the ASE are derived\textcolor{black}{,
and w}e find that there exists an optimal antenna downtilt to achieve
the maximal coverage probability for each base station density. Moreover,
we derive numerically solvable expressions for the optimal antenna
downtilt, which is a function of the base station density. \textcolor{black}{Our
theoretical and numerical results show that after applying the optimal
antenna downtilt, the network performance can be improved significantly.
Specifically, with the optimal antenna downtilt, the ASE crash can
be delayed by nearly one order of magnitude in terms of the base station
density.}
\end{abstract}


\section{Introduction\label{sec:Introduction}}

\begin{comment}
Placeholder
\end{comment}

It has been widely acknowledged that wireless networks continue to
face significant challenges and opportunities. From 1950 to 2000,
the wireless network capacity has increased around 1 million fold~\cite{Webb_survey}.
In the first decade of 2000, network densification continued to fuel
the 3rd Generation Partnership Project (3GPP) 4th-generation (4G)
Long Term Evolution (LTE) networks, and is expected to remain as one
of the main forces to drive the 5th-generation (5G) networks onward~\cite{lopez2015towards}.
Various emerging technologies have been used in cellular networks,
such as small cell networks (SCNs), ultra-dense networks (UDNs), cognitive
radio, massive MIMO, etc~\cite{7422408}. In particular, in the past
few years, a few noteworthy studies have been carried out to revisit\textcolor{black}{{}
the performance analyses} for cellular networks under more practical
propagation assumptions. In~\cite{7061455}, the authors considered
a multi-slope piece-wise path loss function, while in~\cite{6932503},
the authors investigated line-of-sight (LoS) and non-line-of-sight
(NLoS) transmission as a probabilistic event for a millimeter wave
communication scenario. The most important finding in these two works
is that the per-BS coverage probability performance starts to decrease
when the base station (BS) density is sufficiently large. Fortunately,
such decrease of the coverage probability will not change the monotonic
increase of the area spectral efficiency (ASE) as the BS density increases~\cite{6932503,7061455}.
However, in very recent works, the authors found that the ASE performance
will continuously decrease toward zero with the network densification
for UDNs when the absolute height difference betwee\textcolor{black}{n
a ba}se station antenna and \textcolor{black}{a} user equipment (UE)
antenna is larger than zero, which is referred as the ASE Crash in~\cite{8057291,our_work_TWC2016,Iatzeni}. 

Having a closer look at the problem, we realize that in a three-dimensional
(3D) channel model, the antenna pattern and downtilt \textcolor{black}{may}
bring a gain to received signal and at the same time reduce inter-cell
interference~\cite{6839950}. The benefits of horizontal beamforming
in cellular networks are well-understood and such technology has already
been adopted in the LTE networks. However, vertical beamforming (based
on an antenna downtilt) receives much less attention. Recent studies
have made some initial efforts in shedding new light on the impact
of antenna downtilt \textcolor{black}{on} the cellular network~\cite{TR36.814,6692453,7417487},
but most of these studies were solely based on computer simulations.

In this paper, we investigate the impact of the antenna pattern and
downtilt on \textcolor{black}{the performance} of the downlink (DL)
cellular networks, in terms of the coverage probability and the area
spectral efficiency. We also derive the analytical expressions for
the optimal antenna downtilt that resulting in the best coverage probability
of the network given a certain BS density. 

Compared with the existing works, the main contributions of this paper
are: 
\begin{itemize}
\item We analytically investigate the relationship between the the antenna
downtilt and the cellular network performance in terms of the coverage
probability and the ASE. From our theoretical results, we find that
there is a tradeoff between increasing the received signal and reducing
the interference, and hence there exists an optimal antenna downtilt
to achieve the maximal coverage probability for each BS density.
\item We derive numerically solvable expressions for the optimal antenna
downtilt with a certain BS density. In particular, there are three
components, namely the LoS part, the NLoS part and the noise part,
leading to the optimal antenna downtilt. Moreover, we provide analytical
results of the coverage probability and the ASE assuming the optimal
antenna downtilt.
\item Our \textcolor{black}{theoretical and numerical }results demonstrate
that the performance of the cellular network can be improved significantly
using the optimal antenna downtilt. In particular, applying the optimal
antenna downtilt can delay the ASE crash by nearly one order of magnitude
in terms of the base station density. Using the derived expressions
and the simulation results, network operators can determine the antenna
downtilt of BS\textcolor{black}{s} to achieve the optimal system throughput.
\end{itemize}
The rest of this paper is structured as follows. Section II provides
a brief review on the related work. Section III describes the system
model of the 3D cellular network. Section IV presents our theoretical
results on the coverage probability, the optimal antenna downtilt
and the network's performance with the optimal antenna downtilt. The
numerical results are discussed in Section V, with remarks shedding
new light on the network deployment. Finally, the conclusions are
drawn in Section VI.

\section{Related Work\label{sec:Related-Work}}

\begin{comment}
Placeholder
\end{comment}

Stochastic geometry, which is accurate in modeling irregular deployment
of base stations (BSs) and mobile user equipment (UEs), has been widely
used to analyze the network performance~\cite{6042301,6516885}.
Andrews, et al. conducted network performance analyses for the downlink
(DL)~\cite{6042301} and the uplink (UL)~\cite{6516885} of SCNs,
in which UEs and/or BSs were assumed to be randomly deployed according
to a homogeneous Poisson point process (HPPP). Furthermore, a stochastic
model of the 3D environment was used to evaluate the network performance~\cite{8057291,7842150}.
In~\cite{8057291}, Ming, et al. presented a new finding that if
the absolute height difference between BS antenna and UE antenna is
larger than zero, then the ASE performance will continuously decrease
toward zero with the network densification for UDNs. 

Many researchers have realized that a practical antenna can target
its antenna beam towards a given direction via downtilt in the vertical
domain, which may effect the network performance~\cite{6181219,6239994,6839950}.
For example, the authors in~\cite{6839950} found that the antenna
downtilt could bring a significant improvement to the \textcolor{black}{cellular}
network capacity via computer simulations. In~\cite{6181219}, the
authors showed that the vertical beamforming could increase SIR by
about 5-10 dB for a set of UE locations. N. Seifi and M. Coldrey investigated
the performance impact of using antenna downtilt in traditional hexagonal
3D cellular networks in~\cite{6239994}. As we can see, most of the
works that investigated the effect of the antenna downtilt using field
trials or simulations. To the best of our knowledge, none of the existing
works have theoretically analyzed the impact of the antenna downtilt
of BSs on the cellular network performance . 

In this work, we will investigate the impact of the antenna pattern
and downtilt on \textcolor{black}{the performance} of the downlink
(DL) cellular networks and derive the analytical expressions for the
optimal antenna downtilt to achieve the best coverage probability
of the network for each certain BS density. 

\section{System Model\label{sec:System-Model}}

In this section, we will first explain the scenario of the 3D random
cellular network. Then, we will present the antenna patterns and user
association scheme used in this work.

\subsection{Scenario Description\label{subsec:Scenario-Description}}

\begin{comment}
Placeholder
\end{comment}

We consider a 3D random cellular network with downlink (DL) transmissions,
where BSs are deployed on a plane according to an HPPP $\Phi$ of
intensity $\lambda_{B}$ BSs/km$^{2}$. UEs are also Poissonly distributed
in the considered area with an intensity of $\lambda^{\mathtt{UE}}$
UEs/km$^{2}$. Note that $\lambda^{\mathtt{UE}}$ is assumed to be
sufficiently larger than $\lambda_{B}$ so that each BS has at least
one associated UE in its coverage~\cite{6042301,Ding2017capScaling,our_GC_paper_2015_HPPP}.
The two-dimensional (2D) distance between an arbitrary BS and an arbitrary
UE is denoted by $r$ in $m$. Moreover, the absolute antenna height
difference between a BS and a UE is denoted by $L$. Note that the
value of $L$ is in the order of several meters. Hence, the 3D distance
$w$ between a BS and a UE can be expressed as
\begin{equation}
w=\sqrt{r^{2}+L^{2}},
\end{equation}
where $L=H-h$ , $H$ is the antenna height of BS and $h$ is the
antenna height of UE. Intuitively, the antenna height of BS should
decrease as the network becomes dense, however, there is no consensus
on the formula about how $H$ should decrease with an increase in
$\lambda_{B}$. In this paper, we assume that $H$, and thus $L$,
are constants. For the current 4G networks, $L$ is around $8.5m$
because the BS antenna height and the UE antenna height are assumed
to be $10m$ and $1.5m$, respectively~\cite{TR36.814}.

In addition, we incorporate both NLoS and LoS transmissions into the
path loss model. Following~\cite{our_GC_paper_2015_HPPP,our_work_TWC2016},
we adopt a very general path loss model, in which the path loss $\zeta\left(w\right)$,
as a function of the distance $r$, is segmented into $N$ pieces
written as%
\begin{comment}
\begin{singlespace}
\noindent 
\[
\zeta\left(w\right)=\begin{cases}
\zeta_{1}\left(w\right)=\begin{cases}
\begin{array}{l}
\zeta_{1}^{\textrm{L}}\left(w\right),\\
\zeta_{1}^{\textrm{NL}}\left(w\right),
\end{array} & \hspace{-0.3cm}\begin{array}{l}
\textrm{with probability }\textrm{Pr}_{1}^{\textrm{L}}\left(w\right)\\
\textrm{with probability }\left(1-\textrm{Pr}_{1}^{\textrm{L}}\left(w\right)\right)
\end{array}\end{cases}\hspace{-0.3cm}, & \hspace{-0.3cm}\textrm{when }0\leq w\leq d_{1}\\
\zeta_{2}\left(w\right)=\begin{cases}
\begin{array}{l}
\zeta_{2}^{\textrm{L}}\left(w\right),\\
\zeta_{2}^{\textrm{NL}}\left(w\right),
\end{array} & \hspace{-0.3cm}\begin{array}{l}
\textrm{with probability }\textrm{Pr}_{2}^{\textrm{L}}\left(w\right)\\
\textrm{with probability }\left(1-\textrm{Pr}_{2}^{\textrm{L}}\left(w\right)\right)
\end{array}\end{cases}\hspace{-0.3cm}, & \hspace{-0.3cm}\textrm{when }d_{1}<w\leq d_{2}\\
\vdots & \vdots\\
\zeta_{N}\left(w\right)=\begin{cases}
\begin{array}{l}
\zeta_{N}^{\textrm{L}}\left(w\right),\\
\zeta_{N}^{\textrm{NL}}\left(w\right),
\end{array} & \hspace{-0.3cm}\begin{array}{l}
\textrm{with probability }\textrm{Pr}_{N}^{\textrm{L}}\left(w\right)\\
\textrm{with probability }\left(1-\textrm{Pr}_{N}^{\textrm{L}}\left(w\right)\right)
\end{array}\end{cases}\hspace{-0.3cm}, & \hspace{-0.3cm}\textrm{when }w>d_{N-1}
\end{cases}.
\]
\end{singlespace}
\end{comment}
\begin{equation}
\zeta\left(w\right)=\begin{cases}
\zeta_{1}\left(w\right), & \textrm{when }0\leq w\leq d_{1}\\
\zeta_{2}\left(w\right), & \textrm{when }d_{1}<w\leq d_{2}\\
\vdots & \vdots\\
\zeta_{N}\left(w\right), & \textrm{when }w>d_{N-1}
\end{cases},\label{eq:prop_PL_model}
\end{equation}
where each piece $\zeta_{n}\left(w\right),n\in\left\{ 1,2,\ldots,N\right\} $
is modeled as
\begin{equation}
\zeta_{n}\left(w\right)\hspace{-0.1cm}=\hspace{-0.1cm}\begin{cases}
\hspace{-0.2cm}\begin{array}{l}
\zeta_{n}^{\textrm{L}}\left(w\right)=A^{{\rm {L}}}w^{-\alpha_{n}^{{\rm {L}}}},\\
\zeta_{n}^{\textrm{NL}}\left(w\right)=A^{{\rm {NL}}}w^{-\alpha_{n}^{{\rm {NL}}}},
\end{array} & \hspace{-0.2cm}\hspace{-0.3cm}\begin{array}{l}
\textrm{LoS Probability:}~\textrm{Pr}_{n}^{\textrm{L}}\left(w\right)\\
\textrm{NLoS Probability:}~1-\textrm{Pr}_{n}^{\textrm{L}}\left(w\right)
\end{array}\hspace{-0.1cm},\end{cases}\label{eq:PL_BS2UE}
\end{equation}
where 
\begin{itemize}
\item $\zeta_{n}^{\textrm{L}}\left(w\right)$ and $\zeta_{n}^{\textrm{NL}}\left(w\right),n\in\left\{ 1,2,\ldots,N\right\} $
are the $n$-th piece path loss functions for the LoS transmission
and the NLoS transmission, respectively,
\item $A^{{\rm {L}}}$ and $A^{{\rm {NL}}}$ are the path losses at a reference
distance $w=1$ for the LoS and the NLoS cases, respectively,
\item $\alpha_{n}^{{\rm {L}}}$ and $\alpha_{n}^{{\rm {NL}}}$ are the path
loss exponents for the LoS and the NLoS cases, respectively.
\end{itemize}
\noindent In practice, $A^{{\rm {L}}}$, $A^{{\rm {NL}}}$, $\alpha_{n}^{{\rm {L}}}$
and $\alpha_{n}^{{\rm {NL}}}$ are constants obtainable from field
tests and continuity constraints~\cite{SCM_pathloss_model}. 

As a special case, we consider a path loss function adopted in the
3GPP~\cite{TR36.828} as

\begin{equation}
\zeta\left(w\right)\hspace{-0.1cm}=\hspace{-0.1cm}\begin{cases}
\hspace{-0.2cm}\begin{array}{l}
A^{{\rm {L}}}w^{-\alpha^{{\rm {L}}}},\\
A^{{\rm {NL}}}w^{-\alpha^{{\rm {NL}}}},
\end{array} & \hspace{-0.2cm}\hspace{-0.3cm}\begin{array}{l}
\textrm{LoS Probability:}~\textrm{Pr}^{\textrm{L}}\left(w\right)\\
\textrm{NLoS Probability:}~1-\textrm{Pr}^{\textrm{L}}\left(w\right)
\end{array}\hspace{-0.1cm},\end{cases}\label{eq:PL_BS2UEspecial case-1}
\end{equation}
together with a linear LoS probability function as follows~\cite{TR36.828},
\begin{equation}
\textrm{Pr}^{\textrm{L}}\left(r\right)=\begin{cases}
1-\frac{w}{d_{1}} & 0<w\leq d_{1}\\
0 & w>d_{1}
\end{cases},\label{eq:LoS probability function-2}
\end{equation}
where $d_{1}$ is the 3D cut-off distance of the LoS link for BS-to-UE
links. The adopted linear LoS probability function is very useful
because it can include other LoS probability functions as its special
cases~\cite{our_work_TWC2016}.

\begin{figure}
\centering{}\includegraphics[width=8.8cm]{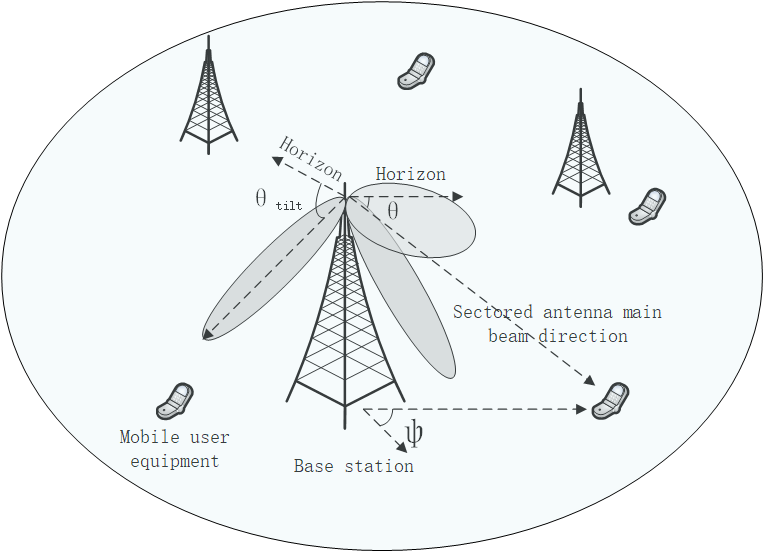}\caption{\label{fig:3D-network}An illustration of the 3D network with randomly
deployed base stations and mobile users}
\end{figure}

Fig.\ref{fig:3D-network} shows an example of the resulted network.
In this model, BSs transmit at power $P_{B}$, a mobile can reliably
communicate with a BS only when its downlink signal-to-interference-plus-noise
ratio (SINR) with respect to that BS is greater than $\gamma$. In
addition, each BS has a 3D directional antenna pattern and we denote
the vertical antenna downtilt and the angle from the BS to the UE
by $\theta_{tilt}$ and $\theta$, respectively.

\subsection{3D Antenna Patterns\label{subsec:3D-Antenna-Patterns}}

3D antenna patterns are introduced in this subsection. According to~\cite{TR36.814}
and~\cite{gunnarsson2008downtilted}, the 3D antenna gain $G(\varphi,\theta,\theta_{tilt})$
can be approximated in dBi as
\begin{equation}
G(\varphi,\theta,\theta_{tilt})=G_{h}(\varphi)+G_{v}(\theta,\theta_{tilt})+G_{m},\label{eq:Gactt}
\end{equation}
where $G_{h}(\varphi)$ and $G_{v}(\theta,\theta_{tilt})$ are the
normalized horizontal and vertical antenna gain in dBi, respectively.
We consider the horizontal omni antenna in this paper, i.e., $G_{h}(\varphi)=0$
dBi. $G_{m}$ is the maximum antenna gain and we can get $G_{m}=8.15dB$
from \cite{TR36.828}. For the vertical pattern, we consider the dipole
antennas. With electrical downtilt~\cite{gunnarsson2008downtilted},
the vertical pattern of the dipole antenna main lobe can be approximated
as
\begin{equation}
G_{v}(\theta,\theta_{tilt})_{dB}=\max\left\{ 10\log_{10}\left|\cos^{n}\left(\theta-\theta_{tilt}\right)\right|,~F_{v2}\right\} ,\label{eq:G}
\end{equation}
 where $F_{v2}=-12dB$ is the vertical side-lobe level. $n=47.64$
for a 4-element half-wave dipole antenna and$\theta=\arctan\left(\frac{L}{r}\right)$
is the angle from the BS to the UE, where $L$ is the height difference
between the BS to the UE, $r$ is the distance from the transmitter
to the receiver. $\theta_{tilt}$ is the vertical antenna downtilt.

\subsection{User Association and Performance Metrics\label{subsec:User-Association}}

In this paper, we assume a practical user association strategy (UAS),
that each UE is connected to the BS with the strongest received power
strength~\cite{our_work_TWC2016,6932503}. Note that in our previous
work~\cite{our_GC_paper_2015_HPPP} and some other existing works,
e.g.,~\cite{6042301,7061455}, it was assumed that each UE should
be associated with its closest BS. Such assumption is not appropriate
for the considered path loss model in Eq.(\ref{eq:prop_PL_model}),
because in practice a UE should connect to a BS offering the largest
received signal strength. Such BS does not necessarily have to be
the nearest one to the UE, and it could be a farther one with a stronger
LoS path. 

Based on the above definitions, we define the coverage probability
as a probability that a receiver's signal-to-interference-plus-noise
ratio (SINR) is above a per-designated threshold $\gamma$:

\noindent 
\begin{equation}
p^{\textrm{cov}}\left(\lambda_{B},\gamma\right)=\textrm{Pr}\left[\mathrm{SINR}>\gamma\right],\label{eq:Coverage_Prob_def}
\end{equation}
where the SINR is calculated as
\begin{equation}
\mathrm{SINR}=\frac{P_{B}G(\varphi,\theta,\theta_{tilt})\zeta\left(r\right)g}{I+N_{0}},\label{eq:SINR defined}
\end{equation}
where $g$ is the channel gain of Rayleigh fading, which is modeled
as an exponential random variable (RV) with the mean of one, and $P_{B}$
and $N_{0}$ are the transmission power of BS and the additive white
Gaussian noise (AWGN) power at each UE, respectively. $I$ is the
cumulative interference given by
\begin{equation}
I=\sum_{i:\,c_{i}\in\Phi\setminus signal}P_{B}G_{i}(\varphi,\theta_{i},\theta_{tilt})\zeta_{i}\left(r\right)g_{i}.
\end{equation}

Furthermore, similar to~\cite{our_work_TWC2016,our_GC_paper_2015_HPPP},
the area spectral efficiency in $\textrm{bps/Hz/k\ensuremath{m^{2}}}$
can be formulated as
\begin{align}
A^{\textrm{ASE}}\left(\lambda_{B},\gamma_{0}\right) & =\lambda\int_{\gamma_{0}}^{\infty}\log_{2}\left(1+x\right)f_{X}\left(\lambda_{B},\gamma_{0}\right)dx,\label{eq:ase}
\end{align}
where $\gamma_{0}$ is the minimum working SINR for the considered
network, and $f_{X}\left(\lambda_{B},\gamma_{0}\right)$ is the probability
density function (PDF) of the SINR observed at the typical receiver
for a particular value of $\lambda$.

\section{Main Results\label{sec:General-Results}}

\begin{comment}
Placeholder
\end{comment}

Using the 3D channel model based on the stochastic geometry theory,
we study the performance of the cellular network and \textcolor{black}{derive}
the optimal antenna downtilt for each certain base station density
in this section. Without any loss of generality we assume that the
mobile user under consideration is located at the origin.

\subsection{The Coverage Probability\label{subsec:The-Coverage-Probability}}

\begin{comment}
Placeholder
\end{comment}

Based on the path loss model in Eq.(\ref{eq:PL_BS2UEspecial case-1})
and the adopted UAS, our results of $p^{\textrm{cov}}\left(\lambda_{B},\gamma\right)$
can be summarized as Theorem~\ref{thm:p_cov_UAS1}, Lemma \ref{lem:-can-be}
and Lemma \ref{lem:-can-be-1}. 
\begin{thm}
\begin{doublespace}
{\small{}\label{thm:p_cov_UAS1}}Considering the path loss model in
Eq.(\ref{eq:PL_BS2UEspecial case-1}) and the presented UAS, the probability
of coverage $p^{{\rm {cov}}}\left(\lambda_{B},\gamma\right)$ can
be derived as
\begin{align}
p^{{\rm {cov}}}\left(\lambda_{B},\gamma\right) & =\int_{0}^{d_{1}}{\rm {Pr}}\left[\left.\frac{\mathtt{S^{L}}}{I_{L}+I_{N}+N_{0}}>\gamma\right|r\right]f_{R,1}^{{\rm {L}}}\left(r\right)dr\nonumber \\
+ & \int_{0}^{d_{1}}{\rm {Pr}}\left[\left.\frac{\mathtt{S^{NL}}}{I_{L}+I_{N}+N_{0}}>\gamma\right|r\right]f_{R,1}^{{\rm {NL}}}\left(r\right)dr\nonumber \\
+ & \int_{d_{1}}^{\infty}{\rm {Pr}}\left[\left.\frac{\mathtt{S^{NL}}}{I_{L}+I_{N}+N_{0}}>\gamma\right|r\right]f_{R,2}^{{\rm {NL}}}\left(r\right)dr\label{eq:coverage}
\end{align}
where $f_{R,1}^{{\rm {L}}}\left(r\right)$ , $f_{R,1}^{{\rm {NL}}}\left(r\right)$
and $f_{R,2}^{{\rm {NL}}}\left(r\right)$ are represented by
\begin{align}
f_{R,1}^{{\rm {L}}}\left(r\right) & =\exp\left(\hspace{-0.1cm}-\hspace{-0.1cm}\int_{0}^{r_{1}}\left(1-{\rm {Pr}}^{{\rm {L}}}\left(u\right)\right)2\pi u\lambda_{B}du\right)\exp\left(\hspace{-0.1cm}-\hspace{-0.1cm}\int_{0}^{r}{\rm {Pr}}^{{\rm {L}}}\left(u\right)2\pi u\lambda_{B}du\right)\nonumber \\
 & {\rm {Pr}}_{1}^{{\rm {L}}}\left(r\right)2\pi r\lambda_{B}\hspace{-0.1cm}\hspace{-0.1cm}\hspace{-0.1cm}\hspace{-0.1cm}
\end{align}
and
\begin{align}
f_{R,1}^{{\rm {NL}}}\left(r\right) & =\exp\left(\hspace{-0.1cm}-\hspace{-0.1cm}\int_{0}^{r_{2}}{\rm {Pr}}^{{\rm {L}}}\left(u\right)2\pi u\lambda_{B}du\right)\exp\left(\hspace{-0.1cm}-\hspace{-0.1cm}\int_{0}^{r}\left(1-{\rm {Pr}}^{{\rm {L}}}\left(u\right)\right)2\pi u\lambda_{B}du\right)\nonumber \\
\times & \left(1-{\rm {Pr}}_{1}^{{\rm {L}}}\left(r\right)\right)2\pi r\lambda_{B}\hspace{-0.1cm}\hspace{-0.1cm}\hspace{-0.1cm}\hspace{-0.1cm}
\end{align}
and
\begin{equation}
f_{R,2}^{{\rm {NL}}}\left(r\right)=\exp\left(\hspace{-0.1cm}-\hspace{-0.1cm}\int_{0}^{r_{2}}{\rm {Pr}}^{{\rm {L}}}\left(u\right)2\pi u\lambda_{B}du\right)\exp\left(\hspace{-0.1cm}-\hspace{-0.1cm}\int_{0}^{r}\left(1-{\rm {Pr}}^{{\rm {L}}}\left(u\right)\right)2\pi u\lambda_{B}du\right)2\pi r\lambda_{B},\label{eq:geom_dis_PDF_UAS1_NLoS_thm-1}
\end{equation}
where $r_{1}$ and $r_{2}$ are given implicitly by the following
equations as
\begin{equation}
r_{1}^{2}=\left(\frac{A_{L}}{A_{NL}}\right)^{-\frac{2}{\alpha_{NL}}}\left(r^{2}+L^{2}\right)^{\frac{\alpha_{L}}{\alpha_{NL}}}-L^{2},\label{eq:def_r_1}
\end{equation}
and
\begin{equation}
r_{2}^{2}=\left(\frac{A_{NL}}{A_{L}}\right)^{-\frac{2}{\alpha_{L}}}\left(r^{2}+L^{2}\right)^{\frac{\alpha_{NL}}{\alpha_{L}}}-L^{2}.\label{eq:def_r_2}
\end{equation}
\end{doublespace}
\end{thm}
\begin{IEEEproof}
See Appendix A.
\end{IEEEproof}
Besides, to compute ${\rm {Pr}}\left[\frac{\mathtt{S^{L}}}{I_{L}+I_{N}+N_{0}}>\gamma\right]$
and ${\rm {Pr}}\left[\frac{\mathtt{S^{NL}}}{I_{L}+I_{N}+N_{0}}>\gamma\right]$
in Theorem \ref{thm:p_cov_UAS1}, we propose Lemma \ref{lem:-can-be}
and Lemma \ref{lem:-can-be-1}, respectively. 
\begin{lem}
\textup{\label{lem:-can-be}${\rm {Pr}}\left[\frac{\mathtt{S^{L}}}{I_{L}+I_{N}+N_{0}}>\gamma\right]$
can be calculated by}
\begin{align}
{\rm {Pr}}\left[\frac{\mathtt{S^{L}}}{I_{L}+I_{N}+N_{0}}>\gamma\right] & =\exp\left(\mathrm{-\frac{\gamma N_{0}}{P_{B}G(\varphi,\theta_{r},\theta_{tilt})A^{L}\sqrt{r^{2}+L^{2}}^{-\alpha^{L}}}}\right)\mathscr{L}_{I_{{\rm {agg}}}}\left(s\right)\nonumber \\
= & \exp\left(\mathrm{-\frac{\gamma N_{0}}{P_{B}G(\varphi,\theta_{r},\theta_{tilt})A^{L}\sqrt{r^{2}+L^{2}}^{-\alpha^{L}}}}\right)\nonumber \\
\times & \exp\left(-2\pi\lambda_{B}\int_{r}^{d_{1}}(1-\frac{\sqrt{u^{2}+L^{2}}}{d_{1}})\frac{u}{1+\frac{G(\varphi,\theta_{r},\theta_{tilt})\sqrt{u^{2}+L^{2}}^{\alpha_{L}}}{\gamma G(\varphi,\theta_{u},\theta_{tilt})\sqrt{r^{2}+L^{2}}^{\alpha^{L}}}}du\right)\nonumber \\
\times & \exp\left(-2\pi\lambda_{B}\int_{r_{1}}^{d_{1}}(\frac{\sqrt{u^{2}+L^{2}}}{d_{1}})\frac{u}{1+\frac{G(\varphi,\theta_{r},\theta_{tilt})A^{L}\sqrt{u^{2}+L^{2}}^{\alpha_{NL}}}{\gamma A^{NL}G(\varphi,\theta_{u},\theta_{tilt})\sqrt{r^{2}+L^{2}}^{\alpha^{L}}}}du\right)\nonumber \\
\times & \exp\left(-2\pi\lambda_{B}\int_{d_{1}}^{\infty}\frac{u}{1+\mathrm{\frac{G(\varphi,\theta_{r},\theta_{tilt})A^{L}\sqrt{u^{2}+L^{2}}^{\alpha_{NL}}}{G(\varphi,\theta_{u},\theta_{tilt})A^{NL}\gamma\sqrt{r^{2}+L^{2}}^{\alpha^{L}}}}}du\right).\label{eq:LOS coverage exp}
\end{align}
\end{lem}
\begin{IEEEproof}
See Appendix A. 
\end{IEEEproof}
\begin{lem}
\textup{\label{lem:-can-be-1}${\rm {Pr}}\left[\frac{\mathtt{S^{NL}}}{I_{L}+I_{N}+N_{0}}>\gamma\right]$
can be calculated by}
\begin{align}
{\rm {Pr}}\left[\frac{\mathtt{S^{NL}}}{I_{L}+I_{N}+N_{0}}>\gamma\right] & =\exp\left(\mathrm{-\frac{\gamma N_{0}}{P_{B}G(\varphi,\theta_{r},\theta_{tilt})A^{NL}\sqrt{r^{2}+L^{2}}^{-\alpha^{NL}}}}\right)\mathscr{L}_{I_{{\rm {agg}}}}\left(s\right),\nonumber \\
= & \exp\left(\mathrm{-\frac{\gamma N_{0}}{P_{B}G(\varphi,\theta_{r},\theta_{tilt})A^{NL}\sqrt{r^{2}+L^{2}}^{-\alpha^{NL}}}}\right)\nonumber \\
\times & \exp\left(-2\pi\lambda_{B}\int_{r_{2}}^{d_{1}}(1-\frac{\sqrt{u^{2}+L^{2}}}{d_{1}})\frac{u}{1+\frac{G(\varphi,\theta_{r},\theta_{tilt})A^{NL}\sqrt{u^{2}+L^{2}}^{\alpha_{L}}}{\gamma A^{L}G(\varphi,\theta_{u},\theta_{tilt})\sqrt{r^{2}+L^{2}}^{\alpha^{NL}}}}du\right)\nonumber \\
\times & \exp\left(-2\pi\lambda_{B}\int_{r}^{d_{1}}(\frac{\sqrt{u^{2}+L^{2}}}{d_{1}})\frac{u}{1+\mathrm{\frac{G(\varphi,\theta_{r},\theta_{tilt})\sqrt{u^{2}+L^{2}}^{\alpha_{NL}}}{\gamma G(\varphi,\theta_{u},\theta_{tilt})\sqrt{r^{2}+L^{2}}^{\alpha^{NL}}}}}du\right)\nonumber \\
\times & \exp\left(-2\pi\lambda_{B}\int_{d_{1}}^{\infty}\frac{u}{1+\mathrm{\frac{G(\varphi,\theta_{r},\theta_{tilt})\sqrt{u^{2}+L^{2}}^{\alpha_{NL}}}{\gamma G(\varphi,\theta_{u},\theta_{tilt})\sqrt{r^{2}+L^{2}}^{\alpha^{NL}}}}}du\right)
\end{align}
where $\theta_{r}=\arctan\left(\frac{L}{r}\right)$ and $\theta_{u}=\arctan\left(\frac{L}{u}\right)$.
\end{lem}
\begin{IEEEproof}
See Appendix A.
\end{IEEEproof}
In Theorem \ref{thm:p_cov_UAS1}, $\mathscr{L}_{I_{{\rm {agg}}}}\left(s\right)$
is the Laplace transform of $I_{{\rm {agg}}}$ evaluated at $s$ including
the LoS interference transmission and that for NLoS transmission.
Regarding the computational process to obtain $p^{{\rm {cov}}}\left(\lambda_{B},\gamma\right)$,
three folds of integrals are respectively required. The string variable
Path takes the value of 'L' and 'NL' for the LoS case and the NLoS
case, respectively. 

\subsection{The impact of antenna downtilt on\textcolor{brown}{{} }\textcolor{black}{the
received signal and the interference }}

The antenna pattern and downtilt \textcolor{black}{may} bring a gain
to the received signal power and at the same time reduce inter-cell
interference. In this subsection, we will analytically investigate
the impact of antenna downtilt on the received signal strength and
the interference of the typical UE, respectively.
\begin{lem}
\label{prop:The-ratio-of}The ratio of the received signal strength
of the typical UE with antenna downtilt to that without can be written
as
\begin{align}
\frac{S_{withG}}{S_{withoutG}} & =G(\varphi,\theta_{r},\theta_{tilt})\nonumber \\
\underset{=}{\left(a\right)} & \cos^{n}\left(\arctan\left(\frac{L}{r}\right)-\theta_{tilt}\right)+10^{0.815}\label{eq:siganl gain}
\end{align}
where $L$ is the antenna height difference between a BS and a UE,
$n=47.64$ for a 4-element half-wave dipole antenna\textup{, $r$
}is the average distance from the transmitter to the receiver in the
network. $\left(a\right)$ can be obtained from Eq.(\ref{eq:Gactt})
when the angel difference $\left(\theta-\theta_{tilt}\right)$ is
small.
\end{lem}
\begin{IEEEproof}
From Theorem \ref{thm:p_cov_UAS1}, the received signal strength with
optimal antenna downtilt can be written as
\begin{equation}
S_{withG}=P_{B}G(\varphi,\theta_{r},\theta_{tilt})\zeta\left(w\right)g
\end{equation}
and the received signal strength without optimal antenna downtilt
can be written as 
\begin{equation}
S_{withoutG}=P_{B}\zeta\left(w\right)g
\end{equation}
Plugging these two into Eq.(\ref{eq:siganl gain}), we have Lemma
\ref{prop:The-ratio-of}, which concludes our proof. 
\end{IEEEproof}
Lemma \ref{prop:The-ratio-of} characterizes the impact of the antenna
downtilt on the received signal. Taking $\lambda_{B}=10^{3}$ BSs/km$^{2}$
as an example, when $r=15.8m$ and $\theta_{optimal}=36\textdegree$,
the ratio of the received signal strength with antenna downtilt to
that without is 6.2529.

In the following we will investigate the performance gain achieved
by bringing down the inter-cell interference power. In Lemma \ref{lem:The-coverage-probability},
we derive the coverage probability without the antenna downtilt in
interference. 
\begin{lem}
\label{lem:The-coverage-probability}The coverage probability that
when the interference without the antenna downtilt can be written
as
\begin{align}
p_{withouG}^{{\rm {cov}}}\left(\lambda_{B},\gamma\right) & =\int_{0}^{d_{1}}\exp\left(-2\pi\lambda_{B}\int_{r}^{d_{1}}(1-\frac{\sqrt{u^{2}+L^{2}}}{d_{1}})\frac{u}{1+\frac{G(\varphi,\theta_{r},\theta_{tilt})\sqrt{u^{2}+L^{2}}^{\alpha_{L}}}{\gamma\sqrt{r^{2}+L^{2}}^{\alpha^{L}}}}du\right.\nonumber \\
- & 2\pi\lambda_{B}\int_{r_{1}}^{d_{1}}(\frac{\sqrt{u^{2}+L^{2}}}{d_{1}})\frac{u}{1+\frac{G(\varphi,\theta_{r},\theta_{tilt})A^{L}\sqrt{u^{2}+L^{2}}^{\alpha_{NL}}}{\gamma A^{NL}\sqrt{r^{2}+L^{2}}^{\alpha^{L}}}}du\nonumber \\
- & \left.2\pi\lambda_{B}\int_{d_{1}}^{\infty}\frac{u}{1+\mathrm{\frac{G(\varphi,\theta_{r},\theta_{tilt})A^{L}\sqrt{u^{2}+L^{2}}^{\alpha_{NL}}}{A^{NL}\gamma\sqrt{r^{2}+L^{2}}^{\alpha^{L}}}}}du\right)f_{R,1}^{{\rm {L}}}\left(r\right)dr\nonumber \\
+ & \int_{0}^{d_{1}}\exp\left(-2\pi\lambda_{B}\int_{r_{2}}^{d_{1}}(1-\frac{\sqrt{u^{2}+L^{2}}}{d_{1}})\frac{u}{1+\frac{G(\varphi,\theta_{r},\theta_{tilt})A^{NL}\sqrt{u^{2}+L^{2}}^{\alpha_{L}}}{\gamma A^{L}\sqrt{r^{2}+L^{2}}^{\alpha^{NL}}}}du\right.\nonumber \\
- & 2\pi\lambda_{B}\int_{r}^{d_{1}}(\frac{\sqrt{u^{2}+L^{2}}}{d_{1}})\frac{u}{1+\mathrm{\frac{G(\varphi,\theta_{r},\theta_{tilt})\sqrt{u^{2}+L^{2}}^{\alpha_{NL}}}{\gamma\sqrt{r^{2}+L^{2}}^{\alpha^{NL}}}}}du\nonumber \\
- & \left.2\pi\lambda_{B}\int_{d_{1}}^{\infty}\frac{u}{1+\mathrm{\frac{G(\varphi,\theta_{r},\theta_{tilt})\sqrt{u^{2}+L^{2}}^{\alpha_{NL}}}{\gamma\sqrt{r^{2}+L^{2}}^{\alpha^{NL}}}}}du\right)f_{R,1}^{{\rm {NL}}}\left(r\right)dr\nonumber \\
+ & \int_{d_{1}}^{\infty}\exp\left(-2\pi\lambda_{B}\int_{r_{2}}^{d_{1}}(1-\frac{\sqrt{u^{2}+L^{2}}}{d_{1}})\frac{u}{1+\frac{G(\varphi,\theta_{r},\theta_{tilt})A^{NL}\sqrt{u^{2}+L^{2}}^{\alpha_{L}}}{\gamma A^{L}\sqrt{r^{2}+L^{2}}^{\alpha^{NL}}}}du\right.\nonumber \\
- & 2\pi\lambda_{B}\int_{r}^{d_{1}}(\frac{\sqrt{u^{2}+L^{2}}}{d_{1}})\frac{u}{1+\mathrm{\frac{G(\varphi,\theta_{r},\theta_{tilt})\sqrt{u^{2}+L^{2}}^{\alpha_{NL}}}{\gamma\sqrt{r^{2}+L^{2}}^{\alpha^{NL}}}}}du\nonumber \\
- & \left.2\pi\lambda_{B}\int_{d_{1}}^{\infty}\frac{u}{1+\mathrm{\frac{G(\varphi,\theta_{r},\theta_{tilt})\sqrt{u^{2}+L^{2}}^{\alpha_{NL}}}{\gamma\sqrt{r^{2}+L^{2}}^{\alpha^{NL}}}}}du\right)f_{R,2}^{{\rm {NL}}}\left(r\right)dr\label{eq:SbiI}
\end{align}
where $f_{R,1}^{{\rm {L}}}\left(r\right)$ , $f_{R,1}^{{\rm {NL}}}\left(r\right)$
and $f_{R,2}^{{\rm {NL}}}\left(r\right)$ can be found in Theorem
\ref{thm:p_cov_UAS1}.
\end{lem}
\begin{IEEEproof}
Note that we consider the antenna downtilt in the received signal
of the typical UE and no antenna downtilt in the interference, therefore
we have
\begin{equation}
S_{signal}=P_{B}G(\varphi,\theta_{r},\theta_{tilt})\zeta\left(w\right)g
\end{equation}
and
\begin{equation}
I=\sum_{i:\,c_{i}\in\Phi_{c}\setminus signal}P_{B}\zeta_{i}\left(r\right)g_{i}.
\end{equation}
Plugging these into Theorem \ref{thm:p_cov_UAS1}, we can get Lemma
\ref{lem:The-coverage-probability}, which concludes our proof. 
\end{IEEEproof}
The difference between Theorem \ref{thm:p_cov_UAS1} and Lemma \ref{lem:The-coverage-probability}
lies in the antenna downtilt gain on interference. \textcolor{black}{Lemma
\ref{lem:The-coverage-probability} and Theorem \ref{thm:p_cov_UAS1}
state that when $\lambda_{B}$ is small, in order to get the best
performance, the chosen antenna downtilt $\theta_{tilt}$ approaches
zero, as $\theta_{u}=\arctan\left(\frac{L}{u}\right)$ is also nearly
zero.}\textcolor{brown}{{} }Therefore, the average antenna gain of the
aggregation interference is larger than 1, e.g., $\mathbb{E_{\mathit{\mathrm{u}}}}\left[G(\varphi,\theta_{u},\theta_{tilt})\right]\thickapprox G_{m}>1$,
which showing that the results of Lemma \ref{lem:The-coverage-probability}
is larger than this of Theorem \ref{thm:p_cov_UAS1}. This means that
the antenna downtilt increases both the received signal power and
the interference power while the former one over-weighs the latter
one. On the other hand, when $\lambda_{B}$ is extremely large, e.g.,
in UDNs, $\mathbb{E_{\mathit{\mathrm{u}}}}\left[G(\varphi,\theta_{u},\theta_{tilt})\right]\thickapprox F_{v2}+G_{m}<1$,
hence the result given by Theorem \ref{thm:p_cov_UAS1} is larger
than that in Lemma \ref{lem:The-coverage-probability}, which means
that the antenna downtilt also reduces inter-cell interference. More
numerical results will be given in Section \ref{subsec:The-Reduction-of}.

\subsection{The Optimal Antenna Downtilt\label{subsec:The-Optimal-Antenna}}

\textcolor{black}{Considering the results of $p^{{\rm {cov}}}\left(\lambda_{B},\gamma\right)$
shown in Eq.(\ref{eq:coverage}) , $\theta_{tilt}$ is the only variable
for certain values of $\lambda_{B}$ and $\gamma$. A large antenna
downtilt reduces inter-cell interference power, while at the same
time decreases signal powers for cell edge UEs. On the other hand,
a small antenna downtilt leads to the opposite case. Therefore, different
antenna downtilts achieve different tradeoffs between the signal power
and the interference power, and hence there exists an optimal antenna
downtilt to achieve the maximal coverage probability for each BS density.}\textcolor{brown}{{}
}However, the math derivation is not tractable when using the antenna
model in Eq.(\ref{eq:Gactt}). In the following, we will use Gaussian
approximation to approximate the antenna downtilt gain $G(\varphi,\theta,\theta_{tilt})$
to obtain tractable results first. Then we will present the optimal
antenna downtilt respecting to the BS density, which is summarized
as Theorem \ref{thm:Then-we-let}.

Using the parameters in~\cite{TR36.814} and the curve fitting function
in MATLAB, the antenna downtilt gain $G(\varphi,\theta,\theta_{tilt})$
can be approximated by a Gaussian function as 
\begin{equation}
G(\varphi,\theta,\theta_{tilt})\thickapprox a\exp\text{\ensuremath{\left[-\frac{(\theta-\theta_{tilt})^{2}}{b}\right]}}+c\label{eq:gaussian function}
\end{equation}
where $a=6.208$ , $b=116.64$ and $c=0.4142$.

Thanks to Eq.(\ref{eq:gaussian function}), it is now possible to
calculate the derivative of the 3-fold integral computation in Eq.(\ref{eq:coverage}),
which improves the tractability of our results. In order to get the
optimal antenna downtilt to maximize the coverage probability, we
take the derivative of the coverage probability and find the optimal
point for each BS density.
\begin{thm}
\label{thm:Then-we-let}For a certain BS density $\lambda$, there
exists an optimal antenna downtilt that can maximize the coverage
probability, and the optimal antenna downtilt satisfies the following
equation:\textup{
\begin{equation}
dP_{c}^{L}(\theta_{tilt})+dP_{c}^{NL}(\theta_{tilt})+dP_{c}^{Noise}(\theta_{tilt})=0\label{eq:optimal}
\end{equation}
where
\begin{align}
dP_{c}^{L}(\theta_{tilt}) & =\int_{0}^{d_{1}}\int_{r}^{\infty}\left(\sqrt{\left(L^{2}+r^{2}\right)}^{\alpha_{L}}\right)\left[(\theta_{v}-\theta_{r})\exp\left[-\frac{\left(\theta_{v}-\theta_{r}\right)\left(\theta_{v}+\theta_{r}-2\theta_{tilt}\right)}{b}\right]\right.\nonumber \\
+ & \left.a\left(\exp\text{\ensuremath{\left[-\frac{(\theta_{v}-\theta_{tilt})^{2}}{b}\right](\theta_{v}-\theta_{tilt})}}-\exp\text{\ensuremath{\left[-\frac{(\theta_{r}-\theta_{tilt})^{2}}{b}\right]\left(\theta_{r}-\theta_{tilt}\right)}}\right)\right]\nonumber \\
\times & f_{R,1}^{{\rm {L}}}\left(r\right)dvdr\label{eq:LOS}
\end{align}
and
\begin{align}
dP_{c}^{NL}(\theta_{tilt}) & =\int_{0}^{d_{1}}\int_{r}^{\infty}\left(\sqrt{\left(L^{2}+r^{2}\right)}^{\alpha_{NL}}\right)\left[(\theta_{v}-\theta_{r})\exp\left[-\frac{\left(\theta_{v}-\theta_{r}\right)\left(\theta_{v}+\theta_{r}-2\theta_{tilt}\right)}{b}\right]\right.\nonumber \\
+ & \left.a\left(\exp\text{\ensuremath{\left[-\frac{(\theta_{v}-\theta_{tilt})^{2}}{b}\right](\theta_{v}-\theta_{tilt})}}-\exp\text{\ensuremath{\left[-\frac{(\theta_{r}-\theta_{tilt})^{2}}{b}\right]\left(\theta_{r}-\theta_{tilt}\right)}}\right)\right]\nonumber \\
\times & f_{R,1}^{{\rm {NL}}}\left(r\right)dvdr\nonumber \\
+ & \int_{d_{1}}^{\infty}\int_{r}^{\infty}\left(\sqrt{\left(L^{2}+r^{2}\right)}^{\alpha_{NL}}\right)\left[(\theta_{v}-\theta_{r})\exp\left[-\frac{\left(\theta_{v}-\theta_{r}\right)\left(\theta_{v}+\theta_{r}-2\theta_{tilt}\right)}{b}\right]\right.\nonumber \\
+ & \left.a\left(\exp\text{\ensuremath{\left[-\frac{(\theta_{v}-\theta_{tilt})^{2}}{b}\right](\theta_{v}-\theta_{tilt})}}-\exp\text{\ensuremath{\left[-\frac{(\theta_{r}-\theta_{tilt})^{2}}{b}\right]\left(\theta_{r}-\theta_{tilt}\right)}}\right)\right]\nonumber \\
\times & f_{R,2}^{{\rm {NL}}}\left(r\right)dvdr\label{eq:NLOS}
\end{align}
and
\begin{align}
dP_{c}^{Noise}(\theta_{tilt}) & =\mathrm{\int_{0}^{d_{1}}\left(\sqrt{\left(L^{2}+r^{2}\right)}^{\alpha_{L}}\right)\frac{aN_{0}\exp\text{\ensuremath{\left[-\frac{(\theta_{r}-\theta_{tilt})^{2}}{b}\right]\frac{2(\theta_{r}-\theta_{tilt})}{b}}}}{P_{B}A_{L}\left(a\exp\text{\ensuremath{\left[-\frac{(\theta_{r}-\theta_{tilt})^{2}}{b}\right]}}+c\right)^{2}}}f_{R,1}^{{\rm {L}}}\left(r\right)dr\nonumber \\
+ & \int_{0}^{d_{1}}\left(\sqrt{\left(L^{2}+r^{2}\right)}^{\alpha_{NL}}\right)\mathrm{\frac{aN_{0}\exp\text{\ensuremath{\left[-\frac{(\theta_{r}-\theta_{tilt})^{2}}{b}\right]\frac{2(\theta_{r}-\theta_{tilt})}{b}}}}{P_{B}A_{NL}\left(a\exp\text{\ensuremath{\left[-\frac{(\theta_{r}-\theta_{tilt})^{2}}{b}\right]}}+c\right)^{2}}}f_{R,1}^{{\rm {NL}}}\left(r\right)dr\nonumber \\
+ & \int_{d_{1}}^{\infty}\left(\sqrt{\left(L^{2}+r^{2}\right)}^{\alpha_{NL}}\right)\mathrm{\frac{aN_{0}\exp\text{\ensuremath{\left[-\frac{(\theta_{r}-\theta_{tilt})^{2}}{b}\right]\frac{2(\theta_{r}-\theta_{tilt})}{b}}}}{P_{B}A_{NL}\left(a\exp\text{\ensuremath{\left[-\frac{(\theta_{r}-\theta_{tilt})^{2}}{b}\right]}}+c\right)^{2}}}f_{R,2}^{{\rm {NL}}}\left(r\right)dr\label{eq:NOISE}
\end{align}
}where $f_{R,1}^{{\rm {L}}}\left(r\right)$ , $f_{R,1}^{{\rm {NL}}}\left(r\right)$
and $f_{R,2}^{{\rm {NL}}}\left(r\right)$ can be found in Theorem
\ref{thm:p_cov_UAS1}\textup{ and }$a=6.208$ , $b=116.64$, $c=0.4142$.
\end{thm}
\begin{IEEEproof}
See Appendix B.
\end{IEEEproof}
From Theorem \ref{thm:Then-we-let}, we can draw the following insights:
\begin{itemize}
\item There are three components in Eq.(\ref{eq:optimal}) which lead to
the optimal antenna downtilt, including the LoS part shown as Eq.(\ref{eq:LOS}),
the NLoS part shown as Eq.(\ref{eq:NLOS}) and the noise part shown
as Eq.(\ref{eq:NOISE}), respectively. 
\item When the networks are sparse, the signal is mostly NLoS and the noise
is the dominant factor. Therefore, the NLoS and noise parts of Eq.(\ref{eq:optimal})
are the major ones that determine the optimal downtilt.
\item As the BS density increases, most signals and a part of interference
links transit from NLoS to LoS, and hence, all components in Eq.(\ref{eq:optimal})
should be taken into account.
\item When the BS density is large enough, almost all signals and the major
interference links are LoS, and the noise is very small compared to
the signal or interference. Therefore, the LoS part of Eq.(\ref{eq:optimal})
is the major component that determines the optimal downtilt.
\end{itemize}
Theorem \ref{thm:Then-we-let} is numerically solvable and we can
use the 'fslove' function in MATLAB to obtain the results.

\subsection{The Area Spectral Efficiency\label{subsec:The-Area-Spectral}}

As mentioned in Eq.(\ref{eq:ase}), we investigate the area spectral
efficiency (ASE) performance in $\textrm{bps/Hz/km}^{2}$, which is
defined as 
\begin{equation}
A^{{\rm {ASE}}}\left(\lambda_{B},\gamma_{0}\right)=\lambda_{B}\int_{\gamma_{0}}^{+\infty}\log_{2}\left(1+\gamma\right)f_{\mathit{\Gamma}}\left(\lambda_{B},\gamma\right)d\gamma,\label{eq:ASE_def}
\end{equation}
where $\gamma_{0}$ is the minimum working SINR for the considered
UDNs, and $f_{\mathit{\Gamma}}\left(\lambda,\gamma\right)$ is the
probability density function (PDF) of the SINR observed at the typical
UE at a particular value of $\lambda_{B}$. Based on the definition
of $p^{{\rm {cov}}}\left(\lambda_{B},\gamma\right)$ in Eq.(\ref{eq:Coverage_Prob_def}),
which is the complementary cumulative distribution function (CCDF)
of SINR, $f_{\mathit{\Gamma}}\left(\lambda_{B},\gamma\right)$ can
be computed by 
\begin{equation}
f_{\mathit{\Gamma}}\left(\lambda_{B},\gamma\right)=\frac{\partial\left(1-p^{{\rm {cov}}}\left(\lambda_{B},\gamma\right)\right)}{\partial\gamma}.\label{eq:cond_SINR_PDF}
\end{equation}
where $p^{{\rm {cov}}}\left(\lambda_{B},\gamma\right)$ can be obtained
from Theorem~\ref{thm:p_cov_UAS1}.

\section{Simulation and Discussion\label{sec:Simulation-and-Discussion}}

\begin{comment}
Placeholder
\end{comment}

In this section, numerical results are provided to validate the accuracy
of our analysis and to verify the intuitive performance trends discussed
in Section~\ref{sec:General-Results}. The analytical results are
compared against Monte Carlo simulation results in the coverage probability.
According to~\cite{TR36.814}, we adopt the following parameters
for 3GPP Case 1: $d_{1}=300m$, $\alpha^{L}=2$, $\alpha^{NL}=3.75$,
$A^{L}=10^{-10.38}$, $A^{NL}=10^{-14.54}$, $P_{B}=24dBm$, $P_{N}=-95dBm$
(including a noise figure of 9 dB at the receivers). 

\subsection{Validation of Theorem~\ref{thm:p_cov_UAS1} on the Coverage Probability}

In this subsection, we investigate the coverage probability and validate
the analytical results in Theorem~\ref{thm:p_cov_UAS1} by comparing
with Monte Carlo simulation results.
\begin{figure}[h]
\centering{}\includegraphics[width=10cm]{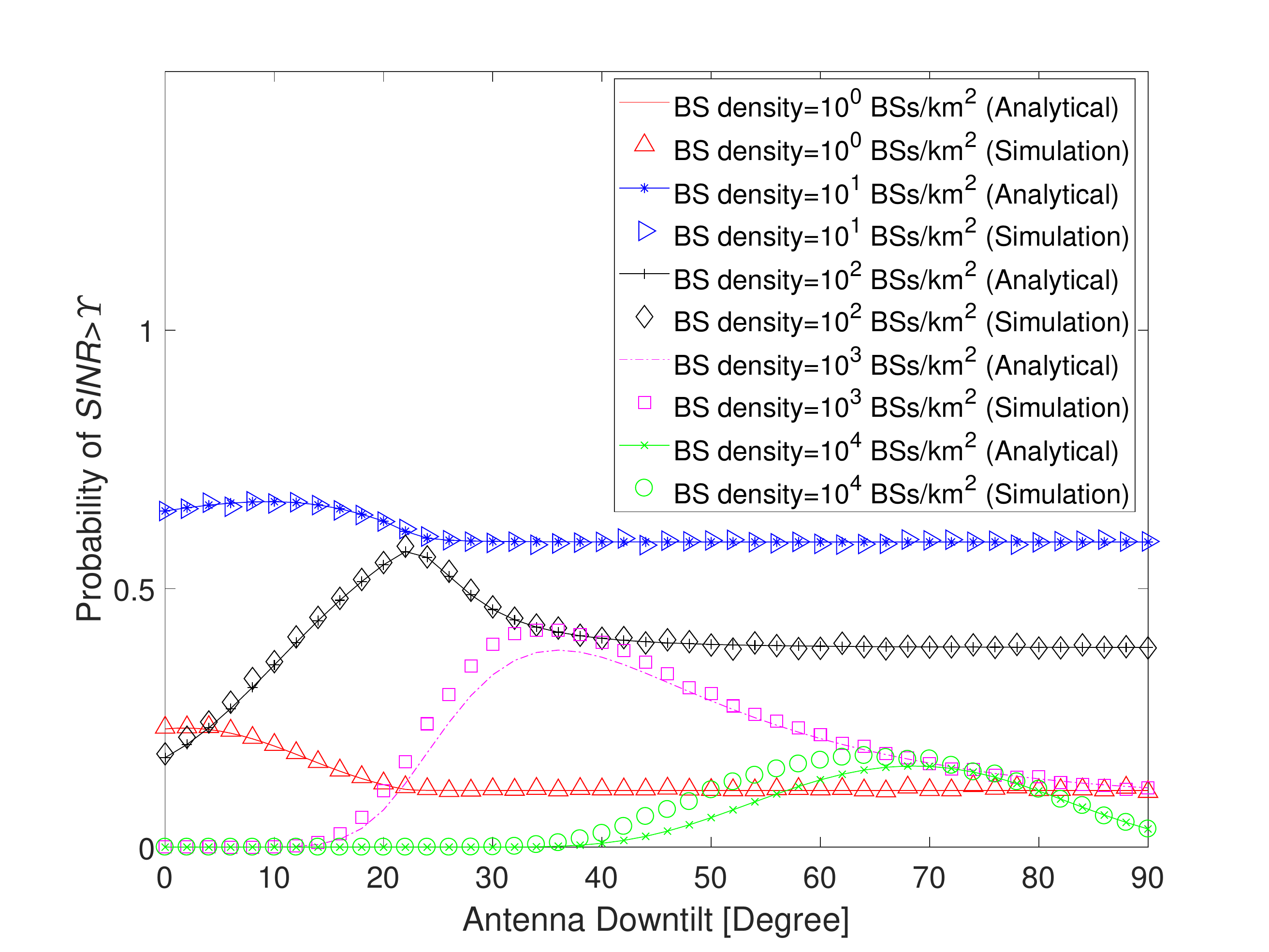}\caption{\label{fig:Coverage-probability-vs.}Coverage probability vs. antenna
downtilt with $\gamma=0dB$ }
\end{figure}
In Fig.\ref{fig:Coverage-probability-vs.}, we plot the results of
the coverage probability for five BS densities with $\gamma=0dB$.
Regarding the non-zero value of $L$, as explained in Section \ref{sec:Introduction},
the BS antenna and the UE antenna heights are set to 10m and 1.5m,
respectively~\cite{TR36.814}. As can be observed from Fig.\ref{fig:Coverage-probability-vs.},
our analytical results given by Theorem \ref{thm:p_cov_UAS1} match
the simulation results very well, and we can draw the following observations:
\begin{itemize}
\item For a certain BS density, there only exists one optimal antenna downtilt
which can achieve the maximum coverage probability. In essence, a
large antenna downtilt reduces inter-cell interference power, while
at the same time decreases signal powers for cell edge UEs. On the
other hand, a small antenna downtilt leads to the opposite case. Therefore,
a different antenna downtilt achieves a different balance between
the signal power and the interference power. 
\item Antenna downtilt has a significant impact on the coverage probability
and the optimal antenna downtilt increases as the BS density increases
from nearly zero degree to 90 degrees.
\end{itemize}

\subsection{Validation of Theorem~\ref{thm:Then-we-let} on the Optimal Antenna
Downtilt}

In this subsection, we validate the analytical results in Theorem~\ref{thm:Then-we-let}
by comparing with Monte Carlo simulation results.
\begin{figure}[H]
\centering{}\includegraphics[width=10cm]{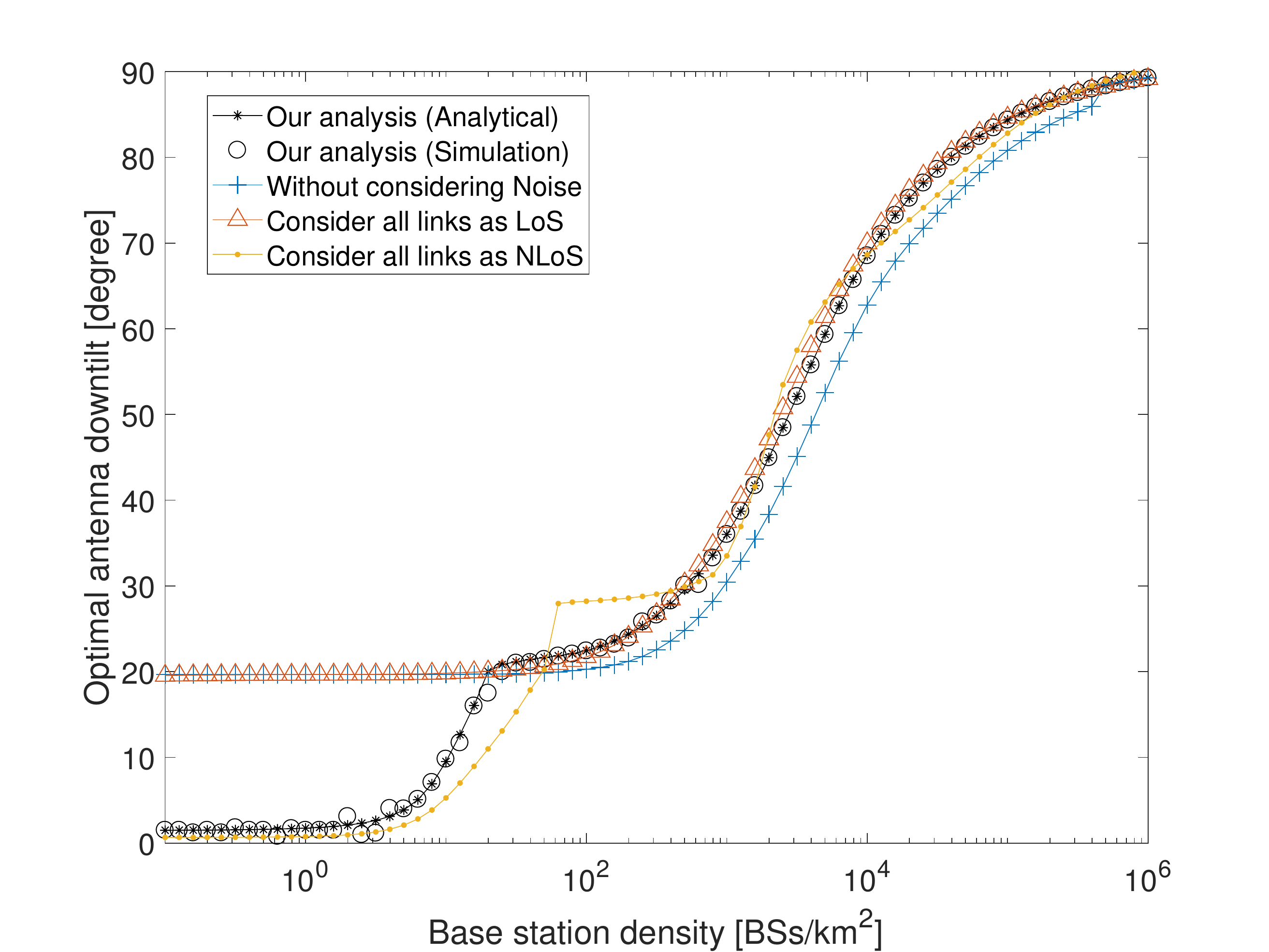}\caption{\label{fig:Coverage-probability-vs.-1}Optimal antenna downtilt vs.
base station density with $\gamma=0dB$ }
\end{figure}

\noindent In Fig.\ref{fig:Coverage-probability-vs.-1}, we show the
optimal downtilt with the BS density increases with $\gamma=0dB$.
As we can observed from Fig.\ref{fig:Coverage-probability-vs.-1},
our analytical results given by Theorem \ref{thm:Then-we-let} match
the simulation results very well, which validates the accuracy of
our analysis. Moreover, we can draw the following observations:
\begin{itemize}
\item The optimal antenna downtilt increases as the BS density increases,
and when the BS density is around $10^{6}$ $BSs/km^{2}$, the optimal
antenna downtilt approaches 90 degree.
\item From Fig.\ref{fig:Coverage-probability-vs.-1}, the curve which considers
all links as NLoS matches results in Theorem \ref{thm:Then-we-let}
when the BSs are sparse. This is due to the fact that the signal is
mostly NLoS and the noise is the dominant factor. Therefore, the NLoS
and noise parts in Eq.(\ref{eq:optimal}) are the major ones that
determine the optimal downtilt.
\item From around $10^{0.3}BSs/km^{2}$ to around $10^{1.1}BSs/km^{2}$,
most signals and interference are NLoS when the BS density is $10^{0.3}BSs/km^{2}$
and then some signals transit from NLoS to LoS \textcolor{black}{and
hence increasing the signal power. The main benefit of the antenna
downtilt is to decrease the dominant interference as the LoS signal
is strong enough.} During this range, both the LoS/NLoS and noise
parts in Eq.(\ref{eq:optimal}) should be considered.
\item When the BS density is around $10^{1.1}BSs/km^{2}$, the increasing
speed of the optimal antenna downtilt is slowing down because most
signals and the dominant interference have transited from NLoS to
LoS, and thus the main purpose of the antenna downtilt shifts from
strengthening the signal power to reducing the interference power.
\item When the BS density is larger than around $10^{1.1}BSs/km^{2}$, almost
all signals are LoS and more and more interference transit from NLoS
to LoS as the BS density increases. The noise is very small compared
to the signal or interference so that the LoS part in Theorem \ref{thm:Then-we-let}
is the major one to determine the optimal antenna downtilt. As we
see from Fig.\ref{fig:Coverage-probability-vs.-1}, when we consider
all links as LoS, the optimal antenna downtilt results are almost
same with the results that achieved based on the model we proposed.
\end{itemize}

\subsection{\label{subsec:The-Gain-of}The Gain of Signal Strength Using Antenna
Downtilt}

\begin{figure}[h]
\centering{}\includegraphics[width=10cm]{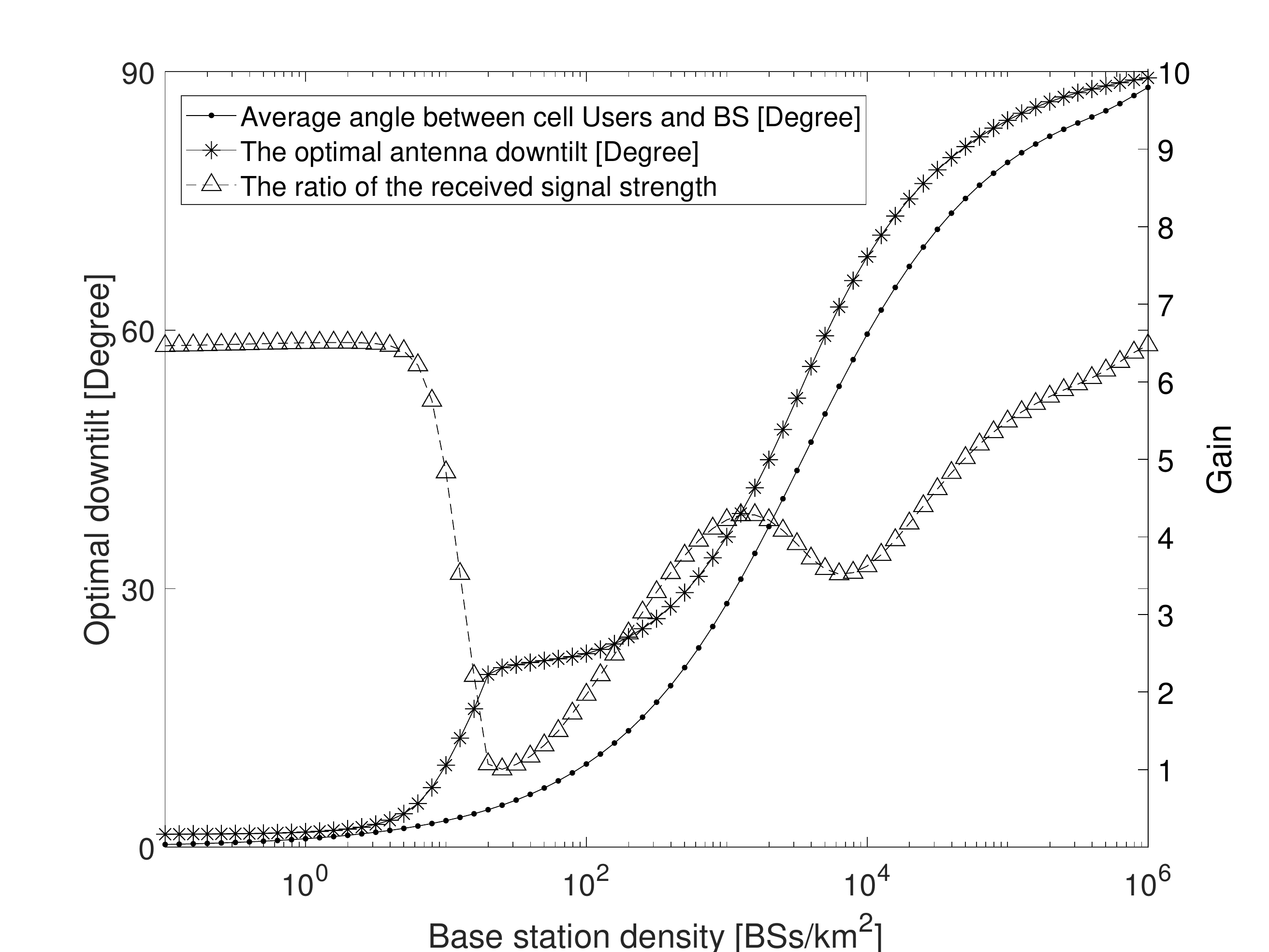}\caption{\label{fig:Signal-gain-vs.}Signal gain vs. base station density }
\end{figure}

Fig.\ref{fig:Signal-gain-vs.} shows the received signal gain of the
typical UE with the BS density which has been analyzed in Eq.(\ref{eq:siganl gain}).
From Fig.\ref{fig:Signal-gain-vs.}, we can see that:
\begin{itemize}
\item When the network is relatively sparse, e.g., from around 0 $BSs/km^{2}$
to around $10^{0.3}BSs/km^{2}$, all links are NLoS. In this case,
through adjusting the antenna downtilt, UE can achieve the maximum
antenna downtilt gain on the received signal, which is around 6.5. 
\item From $10^{0.3}$\textasciitilde{}$10^{1.1}$$BSs/km^{2}$, the received
signal gain brought by the optimal antenna downtilt decreases because
most signal and the dominant interference path from NLoS to LoS, the
main benefit of the antenna downtilt is to decrease the dominant interference. 
\item In the third stage, e.g., from around $10^{1.1}BSs/km^{2}$ to around
$10^{3}BSs/km^{2}$, the received signal gain brought by the optimal
antenna downtilt increases slowly, which means that the increases
of the signal outweighs the decrease of the interference when adopting
the optimal antenna downtilt.
\item Then from around $10^{3}BSs/km^{2}$ to around $10^{3.7}BSs/km^{2}$,
the received signal gain decreases slightly because most interference
transit from NLoS to LoS so that the increase of the aggregation interference
outweighs the decrease of the signal when adopting the optimal antenna
downtilt.
\item In the fifth stage, e.g., from around $10^{3.7}BSs/km^{2}$ to around
$10^{6}BSs/km^{2}$, all links are LoS. In this case, the received
signal gain increases as the BS density increases to obtain the best
coverage probability.
\end{itemize}

\subsection{\label{subsec:The-Reduction-of}The Reduction of Interference Using
Antenna Downtilt}

\begin{figure}[h]
\centering{}\includegraphics[width=10cm]{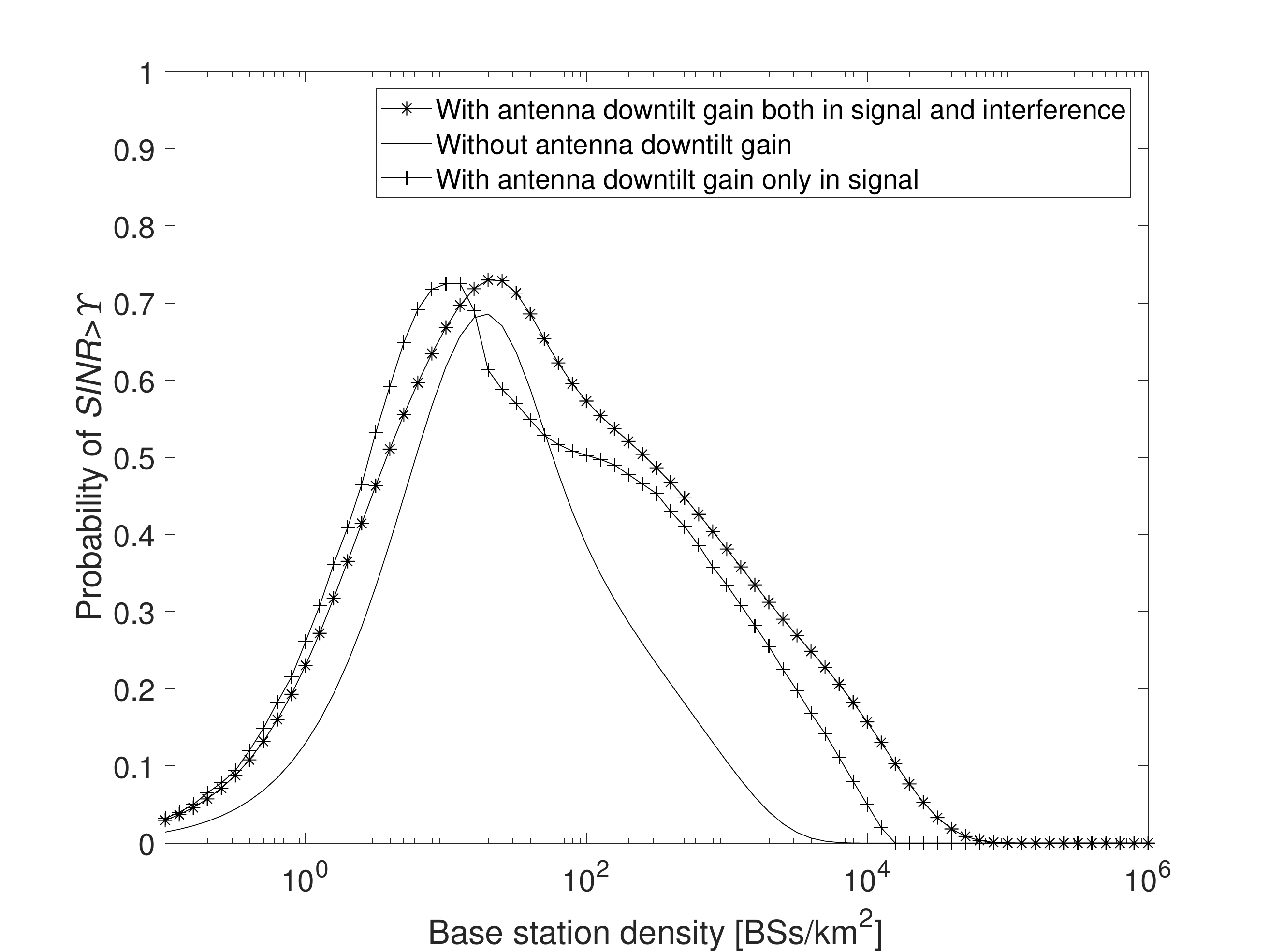}\caption{\label{fig:The-impact-of}The impact of the optimal antenna downtilt
on the interference}
\end{figure}

Fig.\ref{fig:The-impact-of} shows the coverage probability without
antenna downtilt gain on interference. 
From Fig.\ref{fig:The-impact-of}, we can see that:
\begin{itemize}
\item From $0$\textasciitilde{}$10^{1.1}$$BSs/km^{2}$, the gain on the
received signal brought by the optimal antenna downtilt is smaller
than the gain on the interference, therefore the main purpose of antenna
downtilt is to decrease the interference.
\item After $10^{1.1}$$BSs/km^{2}$, the gain on the received signal brought
by the optimal antenna downtilt is larger than the gain on the interference
due to interference reduction. The optimal antenna downtilt brings
down the interference so that improves the coverage probability.
\end{itemize}

\subsection{Network Performance with the Optimal Network-Wide Antenna Downtilt}

In this subsection, we investigate the coverage probability and the
ASE with the optimal antenna downtilt compared with the results in~\cite{lopez2015towards}.

\subsubsection{The coverage probability with the Optimal Network-Wide Antenna Downtilt}

\begin{figure}[H]
\centering{}\includegraphics[width=10cm]{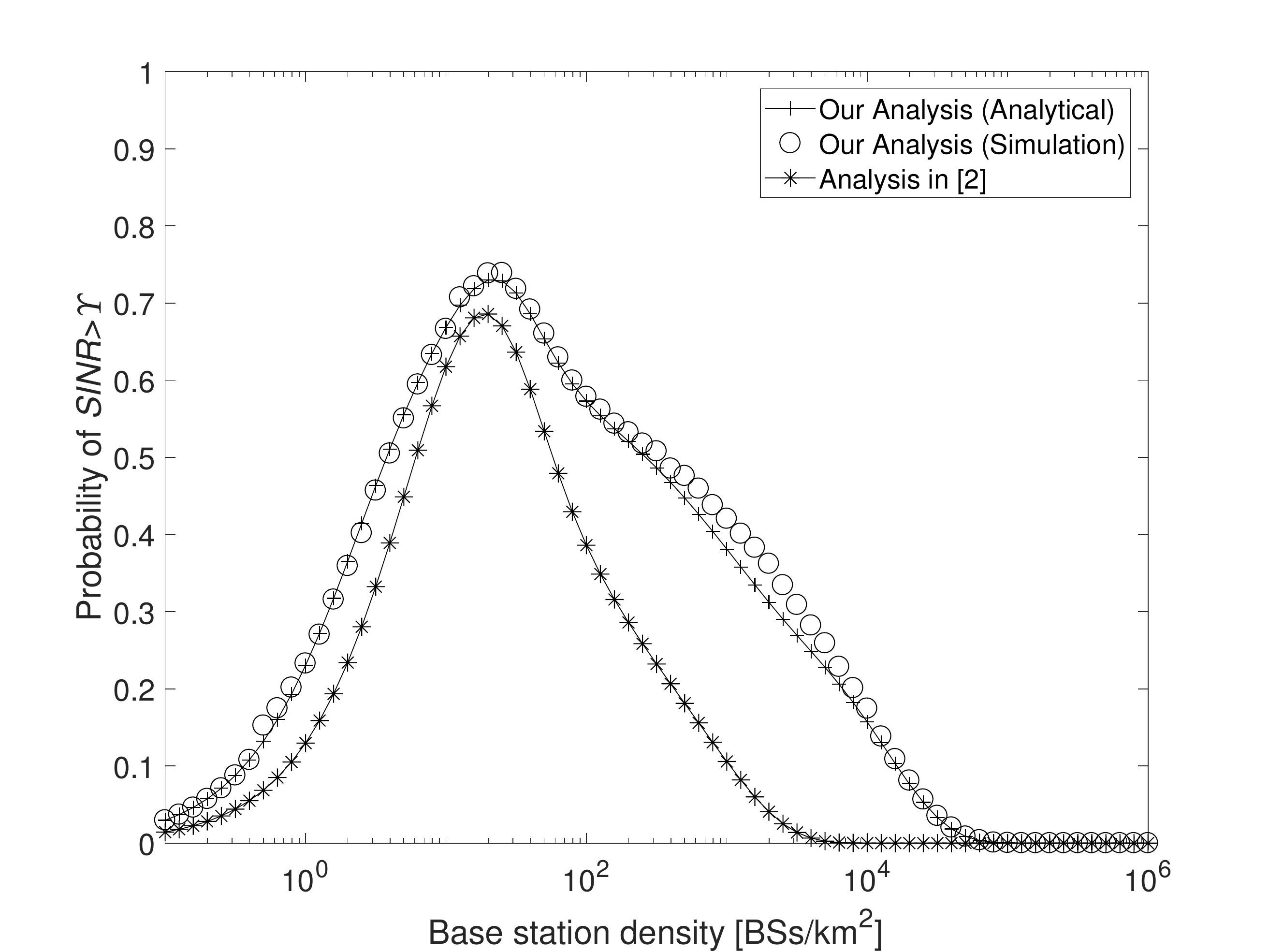}\caption{\label{fig:Coverage-probability-vs.Base}Coverage probability vs.
base station density with optimal antenna downtilt}
\end{figure}
Fig.\ref{fig:Coverage-probability-vs.Base} shows the coverage probability
with the optimal antenna downtilt and without any downtilt. As we
can observe from Fig.\ref{fig:Coverage-probability-vs.Base}:
\begin{itemize}
\item The antenna downtilt does not change the trend of the coverage probability,
i.e., it first increases and then decreases to zero as BS density
increases.
\item The coverage probability performance with the optimal antenna downtilt
is always better than that without antenna downtilt. The coverage
probability reaches zero when the BS density is $3\times10^{4}BSs/km^{2}$
, while it is around $3\times10^{3}BSs/km^{2}$ in the previous work~\cite{lopez2015towards}.
\item Applying the optimal antenna downtilt decreases the rate of decline
of the coverage probability when the BS density is larger than $100BSs/km^{2}$.
\end{itemize}

\subsubsection{The ASE with the Optimal Network-Wide Antenna Downtilt}

In the following, we investigate the ASE performance with the optimal
antenna downtilt. 

\begin{figure}[h]
\begin{centering}
\includegraphics[width=10cm]{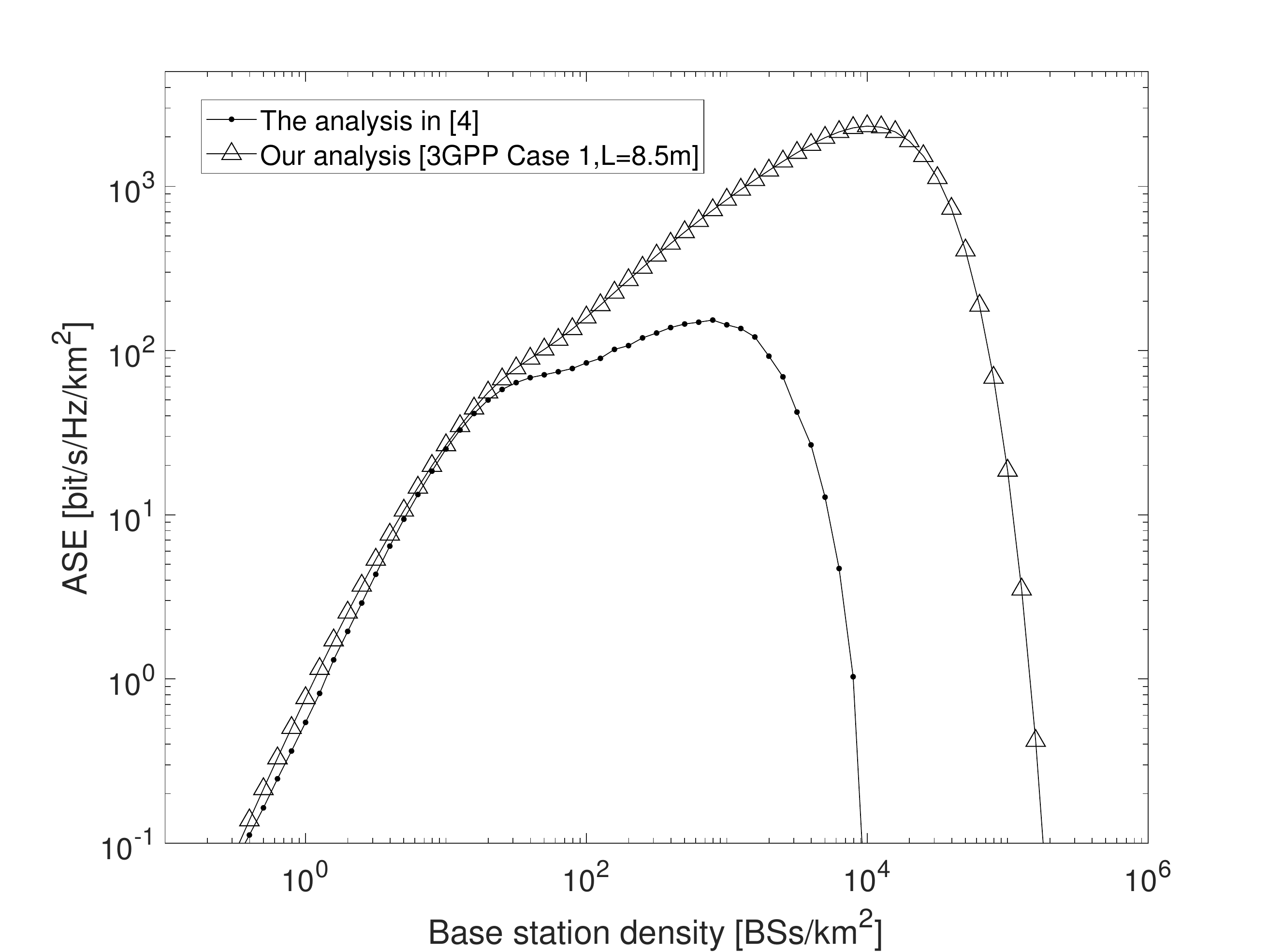}
\par\end{centering}
\caption{\label{fig:-vs.Base-Station's}$A^{ASE}(\lambda,\gamma_{0})$ vs.
base station density with optimal antenna downtilt}
\end{figure}
Fig.\ref{fig:-vs.Base-Station's} shows the ASE with and without optimal
antenna downtilt. From Fig.\ref{fig:-vs.Base-Station's}, we can draw
the following observations:
\begin{itemize}
\item After using the optimal antenna downtilt, the ASE increases as BS
density increases until $2\times10^{4}$ $BSs/km^{2}$, then it decreases
to zero when BS density is around $2\times10^{5}$ $BSs/km^{2}$.
\item The optimal antenna downtilt improves the ASE significantly and delay
the ASE crash by nearly one order of magnitude in terms of the base
station density.
\end{itemize}

\subsection{Network Performance with the Empirical BS-Specific Antenna Downtilt}

In this subsection, we investigate the ASE performance with BS-specific
antenna downtilts. 
\begin{figure}[H]
\begin{centering}
\includegraphics[width=10cm]{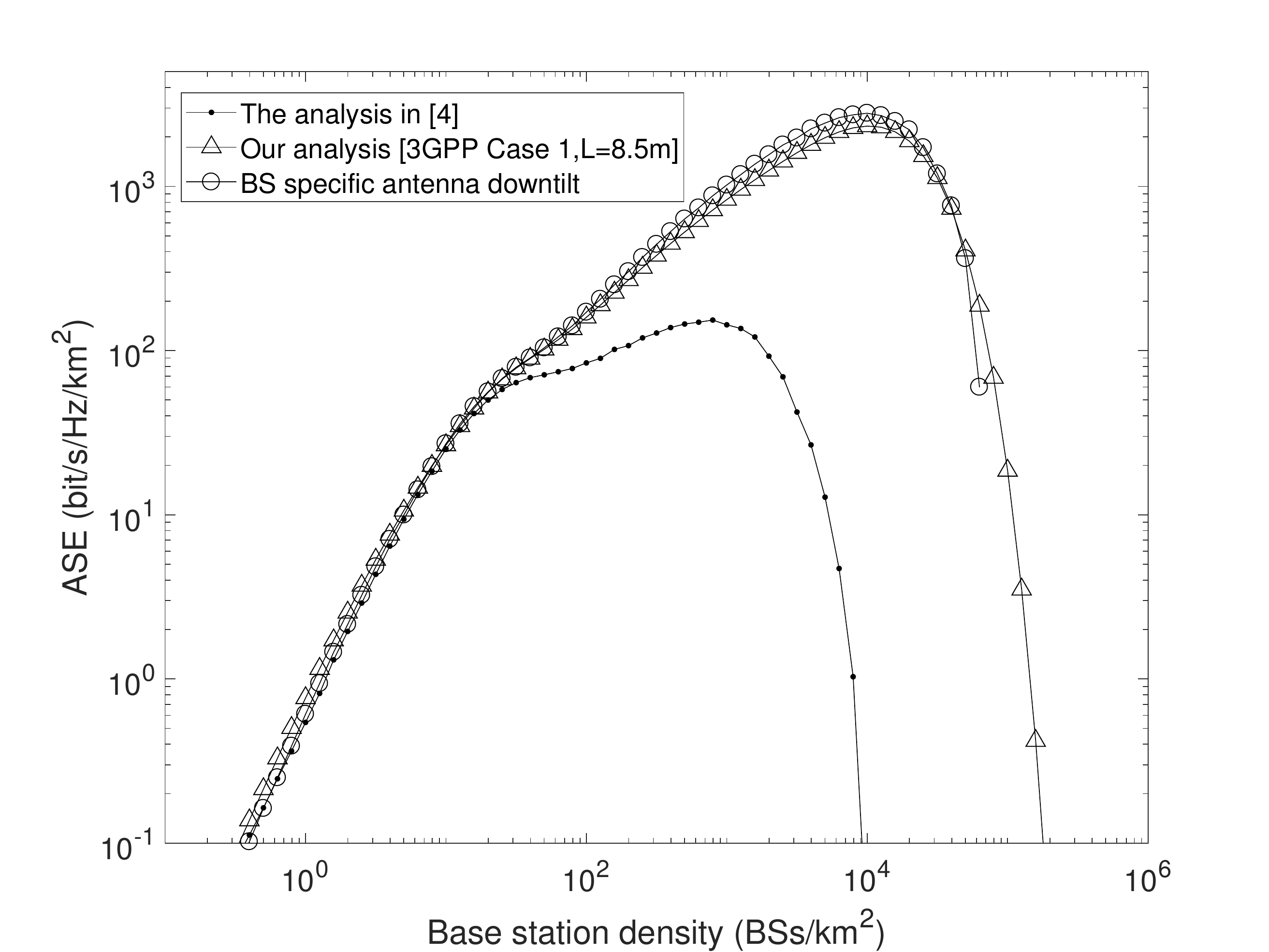}
\par\end{centering}
\caption{\label{fig:-vs.Base-Station's-1}$A^{ASE}(\lambda,\gamma_{0})$ vs.
base station density with optimal antenna downtilt and BS specific
empirical downtilt }
\end{figure}

\begin{figure}
\centering{}\includegraphics[width=6cm]{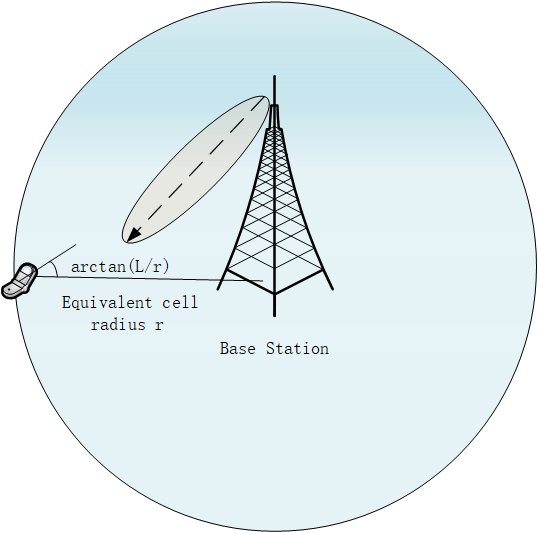}\caption{\label{fig:3D-network-1}An illustrative figure for the empirical
equation}
\end{figure}

In particular, for a certain BS density, adjusting the antenna downtilt
of each base station according to each cell's coverage area may further
improve the performance compared with using an uniform downtilt for
all BSs. For example, for each downlink cell, the BS can adjust its
antenna downtilt based on the distribution of UEs in this particular
cell to maximize the serving signal, instead of using an uniform downtilt
for all cells. In Fig.\ref{fig:-vs.Base-Station's-1}, we investigate
the performance of $A^{ASE}(\lambda,\gamma_{0})$ under the same assumptions
except the choice of antenna downtilt, which uses the BS specific
empiric downtilt. Particularly, each BS adopt an empirical downtilt
as~\cite{lopez2015towards}, which is formulated as
\begin{equation}
\theta_{tilt}=arctam\left(\frac{L}{r}\right)+zB_{V}
\end{equation}
where $r$ is the equivalent radius of each cell, $z$ is set to 0.7
as an empirical value, $B_{V}$ is the vertical half-wave dipole antenna,
for 4-element, $B_{V}=19.5\textdegree$. Fig. \ref{fig:3D-network-1}
illustrates such empirical antenna downtilt.
However, the results showed in Fig.\ref{fig:-vs.Base-Station's-1}
give a sense that the trend of ASE is not changed. From Fig.\ref{fig:-vs.Base-Station's-1},
our key conclusions are drawn as follows:
\begin{itemize}
\item Applying the BS specific empirical antenna downtilt will not change
the trend of ASE as the BS density increases, and the ASE will decrease
towards zero when the BS density is around $2\times10^{5}$ $BSs/km^{2}$.
\item Regarding antenna downtilt, it is not necessary to optimize it on
a per-BS basis as the performance of ASE is not improved much compared
with a network-wide optimal antenna downtilt.
\end{itemize}

\section{Conclusion\label{sec:Conclusion}}

\begin{comment}
Placeholder
\end{comment}

In this paper, we have investigated the impact of the practical antenna
pattern and downtilt on the performance of DL cellular networks. We
\textcolor{black}{found} that there is an optimal antenna downtilt
to achieve the maximal coverage probability for each BS density. Analytical
results have been obtained for the optimal antenna downtilt, the coverage
probability and the ASE performance. Our results have shown that there
are three parts determining the optimal antenna downtilt, and the
optimal antenna downtilt increases as the BS density grows. Compared
with previous works in~\cite{our_work_TWC2016}, we found that using
the optimal antenna downtilt can improve the ASE performance significantly.
Specifically, it can delay the ASE crash by nearly one order of magnitude
in terms of the BS density. As our future work, we will consider the
optimal antenna height in the cellular networks.

\section*{Appendix\ A:Proof of Theorem~\ref{thm:p_cov_UAS1}}
\begin{IEEEproof}
Based on the UAS and the path loss model, the distance of the signal
can be divided into two parts, namely $\left[0,d_{1}\right]$ and
$\left[d_{1},\infty\right]$. In the first path, there are both LoS
and NLoS signal while in the second path, there is only NLoS signal.
For the LoS signal in the first path,
\begin{align}
\Pr\left[\left.\frac{\mathtt{S^{L}}}{I_{L}+I_{N}+N_{0}}>\gamma\right|r\right]\nonumber \\
= & \Pr\left[\left.\frac{P_{B}gG(\varphi,\theta_{r},\theta_{tilt})A^{L}\sqrt{r^{2}+L^{2}}^{-\alpha^{L}}}{I_{L}+I_{N}+N_{0}}>\gamma\right|r\right]\nonumber \\
= & \Pr\left[\left.g>\frac{\gamma\left(I_{L}+I_{N}+N_{0}\right)}{P_{B}G(\varphi,\theta_{r},\theta_{tilt})A^{L}\sqrt{r^{2}+L^{2}}^{-\alpha^{L}}}\right|r\right]\nonumber \\
= & \mathbb{E_{\mathrm{I}}\left[\left.\exp\left(\mathrm{-\frac{\gamma\left(I_{L}+I_{N}+N_{0}\right)}{P_{B}G(\varphi,\theta_{r},\theta_{tilt})A^{L}\sqrt{r^{2}+L^{2}}^{-\alpha^{L}}}}\right)\right|\mathrm{r}\right]}\nonumber \\
= & \exp\left(\mathrm{-\frac{\gamma N_{0}}{P_{B}G(\varphi,\theta_{r},\theta_{tilt})A^{L}\sqrt{r^{2}+L^{2}}^{-\alpha^{L}}}}\right)\mathscr{L}_{I_{{\rm {agg}}}}\left(s\right),
\end{align}
where $s=\mathrm{\frac{\gamma}{P_{B}G(\varphi,\theta_{r},\theta_{tilt})A^{L}\sqrt{r^{2}+L^{2}}^{-\alpha^{L}}}}$,
\begin{align}
\mathscr{L}_{I_{{\rm {agg}}}}\left(s\right) & =\mathbb{E_{\left[\mathit{\mathrm{I}_{r}}\right]}}\left\{ \left.\exp\left(-s\mathit{\mathrm{I}_{r}}\right)\right|0<r<d_{1}\right\} \nonumber \\
= & \mathbb{E_{\left[\mathit{\phi,\left\{ g\right\} ,\left\{ \zeta\left(u\right)G(\varphi,\theta_{u},\theta_{tilt})\right\} }\right]}}\left\{ \left.\exp\left(-sP_{B}g\zeta\left(u\right)G(\varphi,\theta_{u},\theta_{tilt})\right)\right|0<r<d_{1}\right\} \nonumber \\
= & \exp\left(\left.-2\pi\lambda_{B}\int_{r}^{\infty}\left(1-\mathbb{E_{\left[\mathit{\left\{ g\right\} }\right]}}\left\{ \exp\left(-sP_{B}g\zeta\left(u\right)G(\varphi,\theta_{u},\theta_{tilt})\right)\right\} \right)udu\right|0<r<d_{1}\right)\nonumber \\
= & \exp\left(-2\pi\lambda_{B}\int_{r}^{d_{1}}(1-\frac{\sqrt{u^{2}+L^{2}}}{d_{1}})\frac{u}{1+\left(sP_{B}A^{L}G(\varphi,\theta_{u},\theta_{tilt})\right)^{-1}\sqrt{u^{2}+L^{2}}^{\alpha_{L}}}du\right)\nonumber \\
\times & \exp\left(-2\pi\lambda_{B}\int_{r_{1}}^{d_{1}}(\frac{\sqrt{u^{2}+L^{2}}}{d_{1}})\frac{u}{1+\left(sP_{B}A^{NL}G(\varphi,\theta_{u},\theta_{tilt})\right)^{-1}\sqrt{u^{2}+L^{2}}^{\alpha_{NL}}}du\right)\nonumber \\
\times & \exp\left(-2\pi\lambda_{B}\int_{d_{1}}^{\infty}\frac{u}{1+\left(sP_{B}A^{NL}G(\varphi,\theta_{u},\theta_{tilt})\right)^{-1}\sqrt{u^{2}+L^{2}}^{\alpha_{NL}}}du\right)
\end{align}

For the NLoS signal in the first path, in the range of $0<r\leq y_{1}$,
$y_{1}$ means $r_{2}=d_{1}$. 
\begin{align}
\Pr\left[\left.\frac{\mathtt{S^{NL}}}{I_{L}+I_{N}+N_{0}}>\gamma\right|r\right]\nonumber \\
= & \Pr\left[\left.\frac{P_{B}gG(\varphi,\theta_{r},\theta_{tilt})A^{NL}\sqrt{r^{2}+L^{2}}^{-\alpha^{NL}}}{I_{L}+I_{N}+N_{0}}>\gamma\right|r\right]\nonumber \\
= & \Pr\left[\left.g>\frac{\gamma\left(I_{L}+I_{N}+N_{0}\right)}{P_{B}G(\varphi,\theta_{r},\theta_{tilt})A^{NL}\sqrt{r^{2}+L^{2}}^{-\alpha^{NL}}}\right|r\right]\nonumber \\
= & \mathbb{E_{\mathrm{I}}\left[\left.\exp\left(\mathrm{-\frac{\gamma\left(I_{L}+I_{N}+N_{0}\right)}{P_{B}G(\varphi,\theta_{r},\theta_{tilt})A^{NL}\sqrt{r^{2}+L^{2}}^{-\alpha^{NL}}}}\right)\right|\mathrm{r}\right]}\nonumber \\
= & \exp\left(\mathrm{-\frac{\gamma N_{0}}{P_{B}G(\varphi,\theta_{r},\theta_{tilt})A^{NL}\sqrt{r^{2}+L^{2}}^{-\alpha^{NL}}}}\right)\mathscr{L}_{I_{{\rm {agg}}}}\left(s\right),
\end{align}
where $s=\mathrm{\frac{T}{P_{B}G(\varphi,\theta_{r},\theta_{tilt})A^{NL}\sqrt{r^{2}+L^{2}}^{-\alpha^{NL}}}}$,
\begin{align}
\mathscr{L}_{I_{{\rm {agg}}}}\left(s\right) & =\mathbb{E_{\left[\mathit{\mathrm{I}_{r}}\right]}}\left\{ \left.\exp\left(-s\mathit{\mathrm{I}_{r}}\right)\right|0<r<y_{1}\right\} \nonumber \\
= & \mathbb{E_{\left[\mathit{\phi,\left\{ g\right\} ,\left\{ \zeta\left(u\right)G(\varphi,\theta_{u},\theta_{tilt})\right\} }\right]}}\left\{ \left.\exp\left(-sP_{B}g\zeta\left(u\right)G(\varphi,\theta_{u},\theta_{tilt})\right)\right|0<r<y_{1}\right\} \nonumber \\
= & \exp\left(\left.-2\pi\lambda_{B}\int_{r}^{\infty}\left(1-\mathbb{E_{\left[\mathit{\left\{ g\right\} }\right]}}\left\{ \exp\left(-sP_{B}g\zeta\left(u\right)G(\varphi,\theta_{u},\theta_{tilt})\right)\right\} \right)udu\right|0<r<y_{1}\right)\nonumber \\
= & \exp\left(-2\pi\lambda_{B}\int_{r_{2}}^{d_{1}}(1-\frac{\sqrt{u^{2}+L^{2}}}{d_{1}})\frac{u}{1+\left(sP_{B}A^{L}G(\varphi,\theta_{u},\theta_{tilt})\right)^{-1}\sqrt{u^{2}+L^{2}}^{\alpha_{L}}}du\right)\nonumber \\
\times & \exp\left(-2\pi\lambda_{B}\int_{r}^{d_{1}}(\frac{\sqrt{u^{2}+L^{2}}}{d_{1}})\frac{u}{1+\left(sP_{B}A^{NL}G(\varphi,\theta_{u},\theta_{tilt})\right)^{-1}\sqrt{u^{2}+L^{2}}^{\alpha_{NL}}}du\right)\nonumber \\
\times & \exp\left(-2\pi\lambda_{B}\int_{d_{1}}^{\infty}\frac{u}{1+\left(sP_{B}A^{NL}G(\varphi,\theta_{u},\theta_{tilt})\right)^{-1}\sqrt{u^{2}+L^{2}}^{\alpha_{NL}}}du\right)
\end{align}
and when in the range of $y_{1}<r\leq d_{1}$,
\begin{align}
\Pr\left[\left.\frac{\mathtt{S^{NL}}}{I_{N}+N_{0}}>\gamma\right|r\right]\nonumber \\
= & \Pr\left[\left.\frac{P_{B}gG(\varphi,\theta_{r},\theta_{tilt})A^{NL}\sqrt{r^{2}+L^{2}}^{-\alpha^{NL}}}{I_{N}+N_{0}}>\gamma\right|r\right]\nonumber \\
= & \Pr\left[\left.h>\frac{\gamma\left(I_{N}+N_{0}\right)}{P_{B}G(\varphi,\theta_{r},\theta_{tilt})A^{NL}\sqrt{r^{2}+L^{2}}^{-\alpha^{NL}}}\right|r\right]\nonumber \\
= & \mathbb{E_{\mathrm{I}}\left[\left.\exp\left(\mathrm{-\frac{\gamma\left(I_{N}+N_{0}\right)}{P_{B}G(\varphi,\theta_{r},\theta_{tilt})A^{NL}\sqrt{r^{2}+L^{2}}^{-\alpha^{NL}}}}\right)\right|\mathrm{r}\right]}\nonumber \\
= & \exp\left(\mathrm{-\frac{\gamma N_{0}}{P_{B}G(\varphi,\theta_{r},\theta_{tilt})A^{NL}\sqrt{r^{2}+L^{2}}^{-\alpha^{NL}}}}\right)\mathscr{L}_{I_{{\rm {agg}}}}\left(s\right),
\end{align}
where $s=\mathrm{\frac{\gamma}{P_{B}G(\varphi,\theta_{r},\theta_{tilt})A^{NL}\sqrt{r^{2}+L^{2}}^{-\alpha^{NL}}}}$,
\begin{align}
\mathscr{L}_{I_{{\rm {agg}}}}\left(s\right) & =\mathbb{E_{\left[\mathit{\mathrm{I}_{r}}\right]}}\left\{ \left.\exp\left(-s\mathit{\mathrm{I}_{r}}\right)\right|y_{1}<r\leq d_{1}\right\} \nonumber \\
= & \mathbb{E_{\left[\mathit{\phi,\left\{ g\right\} ,\left\{ \zeta\left(u\right)G(\varphi,\theta_{u},\theta_{tilt})\right\} }\right]}}\left\{ \left.\exp\left(-sP_{B}g\zeta\left(u\right)G(\varphi,\theta_{u},\theta_{tilt})\right)\right|y_{1}<r\leq d_{1}\right\} \nonumber \\
= & \exp\left(\left.-2\pi\lambda_{B}\int_{r}^{\infty}\left(1-\mathbb{E_{\left[\mathit{\left\{ g\right\} }\right]}}\left\{ \exp\left(-sP_{B}g\zeta\left(u\right)G(\varphi,\theta_{u},\theta_{tilt})\right)\right\} \right)udu\right|y_{1}<r\leq d_{1}\right)\nonumber \\
= & \exp\left(-2\pi\lambda_{B}\int_{r}^{d_{1}}(\frac{\sqrt{u^{2}+L^{2}}}{d_{1}})\frac{u}{1+\left(sP_{B}A^{NL}G(\varphi,\theta_{u},\theta_{tilt})\right)^{-1}\sqrt{u^{2}+L^{2}}^{\alpha_{NL}}}du\right)\nonumber \\
\times & \exp\left(-2\pi\lambda_{B}\int_{d_{1}}^{\infty}\frac{u}{1+\left(sP_{B}A^{NL}G(\varphi,\theta_{u},\theta_{tilt})\right)^{-1}\sqrt{u^{2}+L^{2}}^{\alpha_{NL}}}du\right)
\end{align}
For the NLoS signal in the second path, in the range of $r>d_{1}$,
\begin{align}
\Pr\left[\left.\frac{\mathtt{S^{NL}}}{I_{N}+N_{0}}>\gamma\right|r\right]\nonumber \\
= & \Pr\left[\left.\frac{P_{B}gG(\varphi,\theta_{r},\theta_{tilt})A^{NL}\sqrt{r^{2}+L^{2}}^{-\alpha^{NL}}}{I_{N}+N_{0}}>\gamma\right|r\right]\nonumber \\
= & \Pr\left[\left.g>\frac{\gamma\left(I_{N}+N_{0}\right)}{P_{B}G(\varphi,\theta_{r},\theta_{tilt})A^{NL}\sqrt{r^{2}+L^{2}}^{-\alpha^{NL}}}\right|r\right]\nonumber \\
= & \mathbb{E_{\mathrm{I}}\left[\left.\exp\left(\mathrm{-\frac{\gamma\left(I_{N}+N_{0}\right)}{PG(\varphi,\theta_{r},\theta_{tilt})A^{NL}\sqrt{r^{2}+L^{2}}^{-\alpha^{NL}}}}\right)\right|\mathrm{r}\right]}\nonumber \\
= & \exp\left(\mathrm{-\frac{\gamma N_{0}}{P_{B}G(\varphi,\theta_{r},\theta_{tilt})A^{NL}\sqrt{r^{2}+L^{2}}^{-\alpha^{NL}}}}\right)\mathscr{L}_{I_{{\rm {agg}}}}\left(s\right),
\end{align}
where $s=\mathrm{\frac{\gamma}{P_{B}G(\varphi,\theta_{r},\theta_{tilt})A^{NL}\sqrt{r^{2}+L^{2}}^{-\alpha^{NL}}}}$,
\begin{align}
\mathscr{L}_{I_{{\rm {agg}}}}\left(s\right) & =\mathbb{E_{\left[\mathit{\mathrm{I}_{r}}\right]}}\left\{ \left.\exp\left(-s\mathit{\mathrm{I}_{r}}\right)\right|r>d_{1}\right\} \nonumber \\
= & \mathbb{E_{\left[\mathit{\phi,\left\{ g\right\} ,\left\{ \zeta\left(u\right)G(\varphi,\theta_{u},\theta_{tilt})\right\} }\right]}}\left\{ \left.\exp\left(-sP_{B}g\zeta\left(u\right)G(\varphi,\theta_{u},\theta_{tilt})\right)\right|r>d_{1}\right\} \nonumber \\
= & \exp\left(\left.-2\pi\lambda_{B}\int_{r}^{\infty}\left(1-\mathbb{E_{\left[\mathit{\left\{ g\right\} }\right]}}\left\{ \exp\left(-sP_{B}g\zeta\left(u\right)G(\varphi,\theta_{u},\theta_{tilt})\right)\right\} \right)udu\right|r>d_{1}\right)\nonumber \\
= & \exp\left(-2\pi\lambda_{B}\int_{d_{1}}^{\infty}\frac{u}{1+\left(sP_{B}A^{NL}G(\varphi,\theta_{u},\theta_{tilt})\right)^{-1}\sqrt{u^{2}+L^{2}}^{\alpha_{NL}}}du\right)
\end{align}
which concludes our proof. 
\end{IEEEproof}

\section*{Appendix B:Proof of Theorem~\ref{thm:Then-we-let}}
\begin{IEEEproof}
\begin{doublespace}
In Theorem~\ref{thm:p_cov_UAS1}
\begin{align}
p^{{\rm {cov}}}\left(\lambda_{B},\gamma\right) & =\int_{0}^{d_{1}}{\rm {Pr}}\left[\left.\frac{\mathtt{S^{L}}}{I_{L}+I_{N}+N_{0}}>\gamma\right|r\right]f_{R,1}^{{\rm {L}}}\left(r\right)dr\nonumber \\
+ & \int_{0}^{d_{1}}{\rm {Pr}}\left[\left.\frac{\mathtt{S^{NL}}}{I_{L}+I_{N}+N_{0}}>\gamma\right|r\right]f_{R,1}^{{\rm {NL}}}\left(r\right)dr\nonumber \\
+ & \int_{d_{1}}^{\infty}{\rm {Pr}}\left[\left.\frac{\mathtt{S^{NL}}}{I_{L}+I_{N}+N_{0}}>\gamma\right|r\right]f_{R,2}^{{\rm {NL}}}\left(r\right)dr.\label{eq:coverage-1}
\end{align}

\end{doublespace}

To get the derivative of $p^{{\rm {cov}}}\left(\lambda_{B},\gamma\right)$
respect to $\theta_{tilt}$, we let $\lambda_{B},\gamma$ be constants.
Except the signal, the other factors which lead to the optimal antenna
downtilt can be divided into the noise part $\varOmega_{noise}$,
the LoS interference part $\Omega_{I_{LoS}}$ and the NLoS interference
parts $\Omega_{I_{NLoS_{1}}}(u<d_{1})$ and $\Omega_{I_{NLoS_{2}}}(u>d_{1})$,
where $u$ is the distance from interference BS to the typical UE.
Then we let the derivative of Eq.(\ref{eq:coverage-1}) be zero, therefore
the three parts in Eq.(\ref{eq:coverage-1}) are all zero. Take the
first part of Eq.(\ref{eq:coverage-1}) as an example, from Eq.(\ref{eq:LOS coverage exp})
\begin{align}
p_{1}^{{\rm {cov}}}\left(\lambda_{B},\gamma\right) & =\int_{0}^{d_{1}}\exp\left\{ \varOmega_{noise}+\Omega_{I_{LoS}}+\Omega_{I_{NLoS_{1}}}+\Omega_{I_{NLoS_{2}}}\right\} f_{R,1}^{{\rm {L}}}\left(r\right)dr
\end{align}
and
\begin{equation}
\int_{0}^{d_{1}}\left\{ \varOmega_{noise}+\Omega_{I_{LoS}}+\Omega_{I_{NLoS_{1}}}+\Omega_{I_{NLoS_{2}}}\right\} _{\theta_{tilt}}'f_{R,1}^{{\rm {L}}}\left(r\right)dr=0\label{eq:devide}
\end{equation}
where
\begin{align}
\varOmega_{noise}+\Omega_{I_{LoS}}+\Omega_{I_{NLoS_{1}}}+\Omega_{I_{NLoS_{2}}}\nonumber \\
= & -2\pi\lambda\int_{r}^{d_{1}}(1-\frac{\sqrt{u^{2}+L^{2}}}{d_{1}})\frac{u}{1+\frac{G(\varphi,\theta_{r},\theta_{tilt})\sqrt{u^{2}+L^{2}}^{\alpha_{L}}}{\gamma G(\varphi,\theta_{u},\theta_{tilt})\sqrt{r^{2}+L^{2}}^{\alpha^{L}}}}du\nonumber \\
- & 2\pi\lambda\int_{r_{1}}^{d_{1}}(\frac{\sqrt{u^{2}+L^{2}}}{d_{1}})\frac{u}{1+\frac{G(\varphi,\theta_{r},\theta_{tilt})A^{L}\sqrt{u^{2}+L^{2}}^{\alpha_{NL}}}{\gamma A^{NL}G(\varphi,\theta_{u},\theta_{tilt})\sqrt{r^{2}+L^{2}}^{\alpha^{L}}}}du\nonumber \\
- & 2\pi\lambda\int_{d_{1}}^{\infty}\frac{u}{1+\mathrm{\frac{G(\varphi,\theta_{r},\theta_{tilt})A^{L}\sqrt{u^{2}+L^{2}}^{\alpha_{NL}}}{G(\varphi,\theta_{u},\theta_{tilt})A^{NL}\gamma\sqrt{r^{2}+L^{2}}^{\alpha^{L}}}}}du\nonumber \\
- & \mathrm{\frac{\gamma N_{0}}{P_{B}G(\varphi,\theta_{r},\theta_{tilt})A^{L}\sqrt{r^{2}+L^{2}}^{-\alpha^{L}}}}
\end{align}
using the Eq.(\ref{eq:gaussian function}), we have 
\begin{equation}
G(\varphi,\theta_{r},\theta_{tilt})'=\frac{2a}{b}(\theta_{r}-\theta_{tilt})\exp\text{\ensuremath{\left[-\frac{(\theta_{r}-\theta_{tilt})^{2}}{b}\right]}}\label{eq:sss}
\end{equation}
and
\begin{equation}
G(\varphi,\theta_{u},\theta_{tilt})'=\frac{2a}{b}(\theta_{u}-\theta_{tilt})\exp\text{\ensuremath{\left[-\frac{(\theta_{u}-\theta_{tilt})^{2}}{b}\right]}}\label{eq:uuu}
\end{equation}
where $\theta_{r}=\arctan\left(\frac{L}{r}\right)$ and $\theta_{u}=\arctan\left(\frac{L}{u}\right)$.
Plugging Eq.(\ref{eq:sss}) and Eq.(\ref{eq:uuu}) into Eq.(\ref{eq:devide}),
and considering all the three parts, we have the Theorem~\ref{thm:Then-we-let},
which concludes our proof.
\end{IEEEproof}
\bibliographystyle{unsrt}
\bibliography{reference}

\begin{thebibliography}{10}

\bibitem{Webb_survey}
{ArrayComm \& William Webb}.
\newblock Ofcom, 2007.

\bibitem{lopez2015towards}
David L{\'o}pez-P{\'e}rez, Ming Ding, Holger Claussen, and Amir~H Jafari.
\newblock Towards 1 {Gbps/UE} in cellular systems: Understanding ultra-dense
  small cell deployments.
\newblock {\em IEEE Communications Surveys \& Tutorials}, 17(4):2078--2101,
  2015.

\bibitem{7422408}
X.~Ge, S.~Tu, G.~Mao, C.~X. Wang, and T.~Han.
\newblock {5G} ultra-dense cellular networks.
\newblock {\em IEEE Wireless Communications}, 23(1):72--79, February 2016.

\bibitem{7061455}
X.~Zhang and J.~G. Andrews.
\newblock Downlink cellular network analysis with multi-slope path loss models.
\newblock {\em IEEE Transactions on Communications}, 63(5):1881--1894, May
  2015.

\bibitem{6932503}
T.~Bai and R.~W. Heath.
\newblock Coverage and rate analysis for millimeter-wave cellular networks.
\newblock {\em IEEE Transactions on Wireless Communications}, 14(2):1100--1114,
  Feb 2015.

\bibitem{8057291}
M.~Ding and D.~L{\'o}pez-P{\'e}rez.
\newblock Performance impact of base station antenna heights in dense cellular
  networks.
\newblock {\em IEEE Transactions on Wireless Communications},
  16(12):8147--8161, Dec 2017.

\bibitem{our_work_TWC2016}
M.~Ding, P.~Wang, D.~López-Pérez, G.~Mao, and Z.~Lin.
\newblock Performance impact of {LoS and NLoS} transmissions in dense cellular
  networks.
\newblock {\em IEEE Transactions on Wireless Communications}, 15(3):2365--2380,
  Mar. 2016.

\bibitem{Iatzeni}
I.~Atzeni, J.~Arnau, and M.~Kountouris.
\newblock Performance analysis of ultra-dense networks with elevated base
  stations.
\newblock In {\em 2017 15th International Symposium on Modeling and
  Optimization in Mobile, Ad Hoc, and Wireless Networks (WiOpt)}, pages 1--6,
  May 2017.

\bibitem{6839950}
A.~D. Gandhi.
\newblock Significant gains in coverage and downlink capacity from optimal
  antenna downtilt for closely-spaced cells in wireless networks.
\newblock In {\em 2014 23rd Wireless and Optical Communication Conference
  (WOCC)}, pages 1--6, May 2014.

\bibitem{TR36.814}
3GPP.
\newblock {TR 36.814: Further advancements for E-UTRA physical layer aspects
  (Release 9)}, Mar. 2010.

\bibitem{6692453}
H.~C. Nguyen, I.~Rodriguez, T.~B. Sorensen, J.~Elling, M.~B. Gentsch,
  M.~Sorensen, and P.~Mogensen.
\newblock Validation of tilt gain under realistic path loss model and network
  scenario.
\newblock In {\em 2013 IEEE 78th Vehicular Technology Conference (VTC Fall)},
  pages 1--5, Sept 2013.

\bibitem{7417487}
J.~Fan, G.~Y. Li, and X.~Zhu.
\newblock Vertical beamforming with downtilt optimization in downlink cellular
  networks.
\newblock In {\em 2015 IEEE Global Communications Conference (GLOBECOM)}, pages
  1--5, Dec 2015.

\bibitem{6042301}
J.~G. Andrews, F.~Baccelli, and R.~K. Ganti.
\newblock A tractable approach to coverage and rate in cellular networks.
\newblock {\em IEEE Transactions on Communications}, 59(11):3122--3134,
  November 2011.

\bibitem{6516885}
T.~D. Novlan, H.~S. Dhillon, and J.~G. Andrews.
\newblock Analytical modeling of uplink cellular networks.
\newblock {\em IEEE Transactions on Wireless Communications}, 12(6):2669--2679,
  June 2013.

\bibitem{7842150}
M.~Ding and D.~L{\'o}pez-P{\'e}rez.
\newblock Please lower small cell antenna heights in 5g.
\newblock In {\em 2016 IEEE Global Communications Conference (GLOBECOM)}, pages
  1--6, Dec 2016.

\bibitem{6181219}
M.~Danneberg, J.~Holfeld, M.~Grieger, M.~Amro, and G.~Fettweis.
\newblock Field trial evaluation of ue specific antenna downtilt in an lte
  downlink.
\newblock In {\em 2012 International ITG Workshop on Smart Antennas (WSA)},
  pages 274--280, March 2012.

\bibitem{6239994}
N.~Seifi, M.~Coldrey, M.~Matthaiou, and M.~Viberg.
\newblock Impact of base station antenna tilt on the performance of network
  mimo systems.
\newblock In {\em 2012 IEEE 75th Vehicular Technology Conference (VTC Spring)},
  pages 1--5, May 2012.

\bibitem{Ding2017capScaling}
M.~{Ding}, D.~L{\'o}pez-P{\'e}rez, and G.~{Mao}.
\newblock A new capacity scaling law in ultra-dense networks.
\newblock {\em arXiv:1704.00399 [cs.NI]}, Apr. 2017.

\bibitem{our_GC_paper_2015_HPPP}
M.~Ding, D.~López-Pérez, G.~Mao, P.~Wang, and Z.~Lin.
\newblock Will the area spectral efficiency monotonically grow as small cells
  go dense?
\newblock {\em IEEE GLOBECOM 2015}, pages 1--7, Dec. 2015.

\bibitem{SCM_pathloss_model}
{Spatial Channel Model AHG}.
\newblock {Subsection 3.5.3, Spatial Channel Model Text Description V6.0}, Apr.
  2003.

\bibitem{TR36.828}
3GPP.
\newblock {TR 36.828: Further enhancements to LTE Time Division Duplex (TDD)
  for Downlink-Uplink (DL-UL) interference management and traffic adaptation},
  Jun. 2012.

\bibitem{gunnarsson2008downtilted}
Fredrik Gunnarsson, Martin~N Johansson, Anders Furuskar, Magnus Lundevall, Arne
  Simonsson, Claes Tidestav, and Mats Blomgren.
\newblock Downtilted base station antennas-a simulation model proposal and
  impact on hspa and lte performance.
\newblock In {\em Vehicular Technology Conference, 2008. VTC 2008-Fall. IEEE
  68th}, pages 1--5. IEEE, 2008.

\end{thebibliography}

\end{document}